\documentclass[aps,prb,twocolumn,showpacs,amsmath,amssymb,superscriptaddress]{revtex4-2}
\usepackage{graphicx}
\usepackage{CJK}
\usepackage{color}
\usepackage[colorlinks,bookmarks=false,citecolor=blue,linkcolor=red,urlcolor=blue]{hyperref}
\usepackage{multirow}
\usepackage{ulem}
\usepackage{tabularx}
\usepackage[nounderscore]{syntax}
\graphicspath{{figs/}}

\newcommand{\be}{\begin{equation}}
\newcommand{\ee}{\end{equation}}

\begin{document}
\title{Extraordinary surface critical behavior induced by symmetry-protected topological state of a two-dimensional quantum magnet}
\author{Zhe Wang}
\affiliation{Department of Physics, Beijing Normal University, Beijing 100875, China}

\author{Fan Zhang}
\affiliation{Department of Physics, Beijing Normal University, Beijing 100875, China}

\author{Wenan Guo}
\email{waguo@bnu.edu.cn}
\affiliation{Department of Physics, Beijing Normal University, Beijing 100875, China}
\affiliation{Beijing Computational Science Research Center, Beijing 100193, China}

\date{\today}
\begin{abstract}
 Using Quantum Monte Carlo simulations, we study spin-1/2 diagonal ladders coupled by ferromagnetic Heisenberg interactions. 
 The model can also be viewed as usual ladders with ferromagnetic rung couplings coupled by antiferromagnetic diagonal couplings.
 We find that the model hosts a striped magnetic ordered phase and two topological nontrivial Haldane phases, separated by two 
 quantum critical points. We show that the two quantum critical points are all in the three-dimensional
 O(3) universality class irrelevant to the topological properties of the Haldane phases.  
 The properties of the surface formed by ladder ends in the two Haldane phases are studied. We find that the surface states are both gapless 
 due to the symmetry-protected topological bulk states. 
 We further demonstrate that extraordinary surface critical behaviors are realized at both critical points on such gapless surfaces 
 without enhancing the surface coupling.  Notably, the surface is not expected to be ordered in the 
 three-dimensional classical O(3) critical point, suggesting that the topological properties of the Haldane phases are responsible 
 for such surface critical behavior. 
\end{abstract}
\maketitle


	\section{Introduction}
	\label{intro}
In the Landau-Ginzburg-Wilson paradigm, phases of matter are characterized by symmetry\cite{landau1980statistical}. 
The universality class (UC) of a critical point separating two phases is determined by the difference of 
symmetries on either side of the critical point and the dimensions of the system\cite{WILSON197475}. 
When a surface presents,  the local environment near the boundary is different from the deep inside the bulk of the system; 
However,  the divergence of correlation length at a critical point makes the influence of boundaries more pronounced; therefore, physical quantities
measured on the surface also show universal behavior, which is called surface critical behavior (SCB)\cite{Cardy}. 

From the classical picture, there are ordinary, special, and extraordinary SCBs \cite{Binder1974, Binder1983, Deng2005}. The singularities of 
the ordinary SCB is purely induced by the bulk criticality when the surface orders simultaneously with the bulk. 
In the case that the surface couplings are sufficiently enhanced, the surface may order 
in advance of the bulk when the temperature is lowered. The ordered surface exhibits extra singularities at the bulk transition 
point. This is called the extraordinary SCB. The special SCB is a multicritical point between the ordinary and the extraordinary transitions. 
For the three-dimensional (3D) model with continuous symmetry, extraordinary SCB does not apply since it's not possible to have an 
ordered two-dimensional (2D) surface.  
Recent research discovered that the surface exhibits an extraordinary-log SCB with sufficiently enhanced surface couplings at the critical point of 
the 3D O($N \geq 2$) UC \cite{Metlitski2020, ParisenToldin2021, Hu2021, ParisenToldin2022, meineri2022, Zou2022}. 
Based on the mapping between a $d$-dimensional quantum system and a $(d + 1)$-dimensional classical system, this general picture of SCBs 
should apply to a quantum critical point (QCP). 

However, distinct topologically ordered quantum phases that cannot be distinguished by symmetry were discovered since the study of the 
fractional quantum Hall effect. One class of such phases has symmetry-protected topological (SPT) order, which cannot be mapped to a product state 
if only symmetric perturbations are allowed \cite{XiaoGang2009, Frank2010, XiaoGang2012}. The symmetries of the perturbations are called the 
protecting symmetries. The spin-1 Haldane chain \cite{Haldane, AKLT} is the first example of such a system. 
SPT phases are often characterized by a gap separating excitations from the ground state in bulk and the presence of gapless or degenerate edge modes.
This connection between bulk and edge properties, known as bulk-edge correspondence, is due to the topology of the state. 

Now consider a critical point separating an SPT phase with a symmetry $G$ and a phase with symmetry spontaneously broken to $H$ in $d$ dimensions.
We may ask the question: Does the topological order affect the UC of the critical point, which, according to the Landau-Ginzburg-Wilson paradigm, 
should be determined by the difference of $G$ and $H$ and the dimensionality $d$? 
Does the surface state of the SPT phase affect the SCBs associated with the critical point? 

These questions have attracted recent investigations\cite{Zhang2017, Zhu2021, Wang2022}. For the models studied previously, it was shown that the 
topological order does not affect the UC of the critical point. 
However,  the gapless edge modes of the SPT phases studied previously, when merging with the bulk critical mode,
lead to ``nonordinary" SCBs at the  ($2+1$)-D O(3) bulk critical point, which is unexpected according to the quantum-classical correspondence. 
Interestingly, nonordinary SCBs characterized by similar critical exponents are also found unexpectedly at the ($2+1$)-D O(3) quantum critical 
point separating topological trivial product states and the symmetry broken phase\cite{Ding2018, Weber2018, Ding2021, Wang2022}, where
the surface formed by the dangling spin chain weakly coupled to the bulk at the product state is also gapless according to the 
Lieb-Schultz-Mattis theorem \cite{LIEB1961}.
The two nonordinary SCBs both have purely quantum but different origins. 

\begin{figure}[htb]
	\centering
	\includegraphics[width=0.5\textwidth]{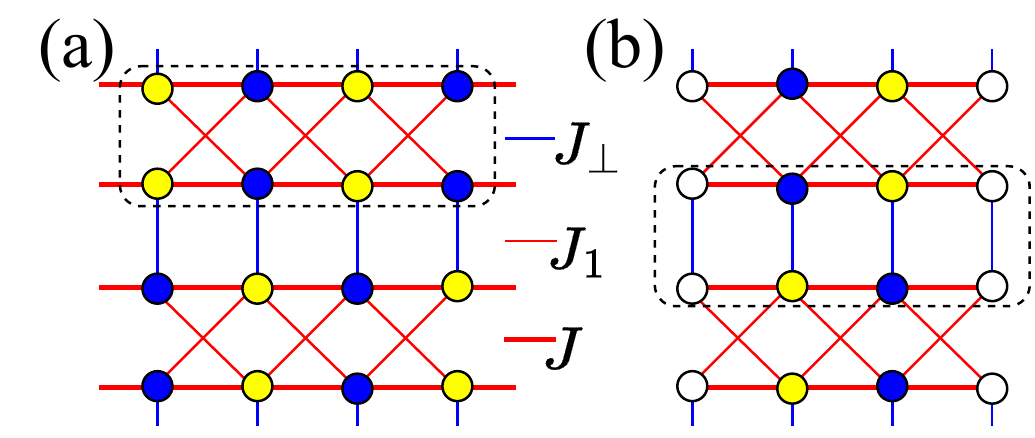}
	\caption{ The two-dimensional coupled diagonal ladders.  
    The lattice is bipartite with sublattices A (yellow circles) and B (blue circles). (a) Periodic
    boundary conditions are applied in $x$ and $y$ directions. A diagonal ladder is shown inside the dashed rectangular box.
    (b) Periodic boundary conditions are applied in $y$ direction, while open boundaries are applied in $x$ direction to expose  surfaces. Open circles denote spins on the surfaces. A usual ladder is shown inside the dashed rectangular box.
    The inter-diagonal-ladder couplings, or equivalently, the rung couplings of the usual ladders,  $J_{\perp}$ are indicated by blue lines, inter-usual-ladder couplings $J_{1}$ by thin red lines, and the couplings $J>0$ by thick red lines. }
	\label{Fig:model}
    \end{figure}

\begin{figure}[htb]
	\centering
	\includegraphics[width=0.5\textwidth]{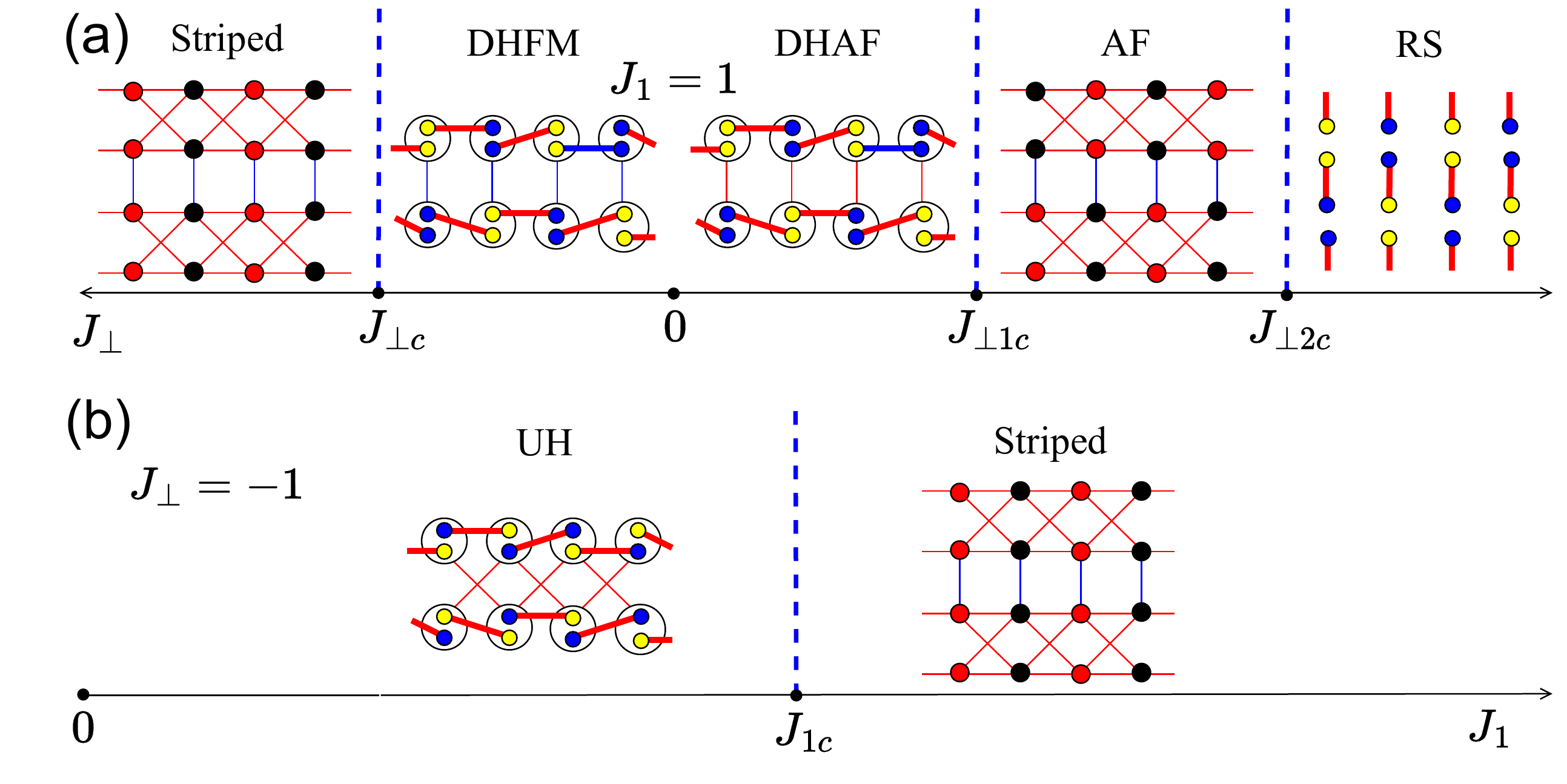}
	\caption{ Phase diagrams for (a)$J_1=1$ and (b) $J_\perp=-1$, respectively. 
 (a) The FM-coupled diagonal-ladder Haldane phase (DHFM) and the  striped  magnetic order phase (Striped)  are separated by the QCP $J_{\perp c}$  in the region of $J_{\perp }<0$. The AF-coupled diagonal-ladder Haldane phase (DHAF), the antiferromagnetic phase (AF), and the rung single phase (RS) are separated by two QCPs $J_{\perp 1c}$ and $J_{\perp 2c}$ in the region of $J_{\perp }>0$. (b) The usual-ladder Haldane phase (UH) and striped  magnetic order phase (Striped) are separated by the QCP $J_{1c}$.  A cartoon of a representative ground state is graphed in each phase. The red and black solid circles in  the striped phase and AF phase represent the orientation of spins. Thick-red lines denote spin singlets.  The circles in the Haldane phases indicate that two spin-1/2 form a spin-1. The $\vert$ and $\times$ in the Haldane phases represent the ways of ladder coupling.  
 }
\label{Fig:diagram}
\end{figure}

In one of these papers \cite{Wang2022},  we studied the coupled diagonal ladders (CDLs) model, which is constructed by coupling 
the spin-1/2 diagonal ladders \cite{diagonal2000} {\it antiferromagnetically} (AF) to form a 2D lattice, as illustrated in 
Fig. \ref{Fig:model}(a) with $J_\perp >0$.   
In nature, there are materials described by quasi-one-dimensional spin-1 chains or spin ladders\cite{Buyers1986, ElbioDagotto1999}. 
 The diagonal ladder is the composite spin representation of a spin-1 Haldane chain, in the sense that the low-energy spectra of the two systems are
 identical\cite{diagonal2000}. 
 When the ladders are weak AF coupled, the ground state is a 2D SPT Haldane phase (DHAF) \cite{Wierschem2014}. 
 We showed that the critical point separating the topological nontrivial phase and the N\'eel phase is in the (2+1)-d O(3) universality class, the 
 same as that separating the topological trivial rung singlet (RS) phase and the N\'eel phase, see Fig. \ref{Fig:diagram}(a).
At the QCP from the SPT Haldane phase to the N\'eel phase, we found nonordinary SCBs on the surface formed by the ends of ladders,
which is attributed to the gapless edge mode of the SPT phase.

The system can also be viewed as usual two-leg ladders with {\it antiferromagnetic} rung couplings $J_\perp$ that are coupled by diagonal 
AF bonds, as illustrated in Fig. \ref{Fig:model}(b). 
Both the diagonal ladder and the usual ladder have short-ranged valence bond (VB) ground states but with different topologies, which are 
defined by the parity of the number 
of VBs crossing an arbitrary line vertical to the ladder. The ground state of the diagonal ladder is odd, which has spin-1/2's localized
at the ends of the ladder for open boundaries, 
but that of the usual ladder with AF rung couplings is even, which does not have spin-1/2's localized at the ends. 
This explains the RS phase does not have a gapless edge mode at the ends of the ladders while the DHAF phase has. 

A natural question arises: what are the properties of the phase that diagonal ladders are {\it ferromagnetically} (FM) coupled, i.e., 
$J_\perp <0$ in Fig. \ref{Fig:model}(a); or, equivalently,  what are the properties of the phase 
that  usual ladders with  FM rung couplings (FM usual ladder with $J_\perp < 0$) are diagonally AF coupled, illustrated in Fig. \ref{Fig:model}(b)?  

Let us start with two limits: first, the FM usual ladders are weakly AF coupled, i.e., $0 < J_1 \ll 1$. 
Note that the usual ladder with FM rungs behaves like a spin-1 chain, and its VB ground state is odd \cite{diagonal2000}. The ground state
should be a 2D SPT Haldane phase \cite{Wierschem2014, Zhu2021} since the couplings between ladders are weak. However, the spins here are 
diagonally coupled,
which may be considered having effective FM couplings between two neighboring spins in different ladders, thus may lead to different properties 
from the DHAF phase, reflected in the edge modes due to the bulk-edge correspondence. This phase is referred to as the usual ladder Haldane (UH) 
phase (See Fig. \ref{Fig:diagram}(b)).
Second, the diagonal ladders are weakly FM coupled, i.e., $|J_\perp| \ll 1$; Again, we expect the ground state to be a 2D SPT Haldane phase
but equivalent to FM coupled spin-1 chains. The FM coupling may lead to different properties from the DHAF phase, 
especially with different edge modes. We refer to this phase as the FM-coupled diagonal ladder Haldane (DHFM) phase (See Fig. \ref{Fig:diagram}(a)).

When the strength of the couplings between $J_1$ and $J_{\perp}$ are comparable, it is natural to expect the system to transfer to a  
magnetically ordered striped phase. We, therefore, expect the phase diagrams sketched in Fig. \ref{Fig:diagram}. 
It is then valuable to check whether the transitions from the two SPT Haldane phases to the striped phase still obey the Landau-Ginzburg-Wilson 
paradigm. Suppose the two SPT Haldane phases are topologically different from the DHAF phase studied in \cite{Wang2022}; we then ask whether the 
new SPTs induce different SCBs at the critical points, considering the bulk-edge correspondence. 

In this work, we answer these questions using unbiased quantum Monte Carlo (QMC) simulations \cite{Sandviksusc1991, Sandvik1999}.
We prove numerically the presence of the magnetically ordered striped phase and the existence of two 
quantum critical points; one is between the striped phase and the SPT DHFM phase, 
and the other is between the striped phase and the SPT UH phase. 
The critical exponents associated with the two QCPs are determined,
showing that they all belong to the (2+1)-D O(3) universality class; this fact reveals that the topological properties of the phases are irrelevant. 
We then show that both SPT phases have gapless edge modes on the surfaces formed by the ends of the ladders.
We further demonstrate that extraordinary SCBs are realized at the two bulk critical points on the gapless surfaces  
without enhancing surface couplings, instead of the nonordinary SCB found at the DHAF and N\'eel QCP in the 
AF-coupled diagonal ladders.
This finding reveals that the SPT UH phase and the SPT DHFM phase are topologically different from the SPT  DHAF phase 
according to the bulk-edge correspondence.

The paper is organized as follows: We describe the model and methods in Sec. \ref{Sec:mm}. Sec. \ref{sec:bulk} presents results of the bulk phase 
transitions and surface properties of the SPT phase.  We show detailed analyses of the surface critical behaviors in Sec. \ref{sec:scb}, then 
conclude in Sec. \ref{sec:conclsn}.

\section{Models and methods}
\label{Sec:mm}

We study the spin-1/2 Heisenberg model on coupled diagonal ladders; see Fig. \ref{Fig:model}(a). The system can also be viewed as  
coupled usual ladders, as shown in Fig. \ref{Fig:model}(b). For convenience, we write the Hamiltonian in terms of coupled diagonal ladders:
 \begin{equation}
	\begin{split}
	H &= \sum_{j=0}H_{j}+ J_{\perp}\sum_{i,j=0}\mathbf{S}_{i,2j+1}\cdot \mathbf{S}_{i,2(j+1)},
		\end{split}
	\end{equation}
	where the first sum is over the diagonal ladders with $H_{j}$ describing the $j$-th ladder written as follows 
	\begin{equation}
	\begin{split}
	H_j &= J \sum_{l=0,1} \sum_i \mathbf{S}_{i,2j+l}\cdot \mathbf{S}_{i+1,2j+l}\\
	&+ J_1\sum_{i}[\mathbf{S}_{i,2j}\cdot \mathbf{S}_{i+1,2j+1}+\mathbf{S}_{i,2j+1}\cdot \mathbf{S}_{i+1,2j}],
		\end{split}
		\label{H_ladder}
	\end{equation}
where $l=0,1$ denote two legs of the $j$-th diagonal ladder, $J>0$ and $J_1>0$  are intra-diagonal-ladder Heisenberg exchange interactions.
The second sum describes the coupling of the neighboring ladders with the inter-diagonal-ladder couplings $J_{\perp}$. 
In terms of coupled usual ladders,  $J$ and $J_{\perp}$ are intra-usual-ladder Heisenberg exchange interactions,  and 
$J_{1}$ is inter-usual-ladder couplings.
In this work, we restrict to the ferromagnetic case $J_{\perp}<0$ and set $J=1$ to fix the energy scale.  

The lattice is bipartite, and the ferromagnetic couplings $J_\perp$ do not introduce magnetic frustration; therefore,  the model can 
be studied using QMC simulations. In this work, we use the stochastic series expansion (SSE) quantum Monte Carlo simulations with the loop update
algorithm \cite{Sandviksusc1991, Sandvik1999} to study the bulk and surface critical behaviors of the model.
Periodic boundary conditions applied in  both   $x$ and $y$ lattice directions are used to study bulk phase transitions.
When the surface states and surface critical behaviors are studied, periodic boundary conditions are applied along the $y$ direction, and open 
boundary conditions are used along the $x$ direction to expose the surfaces, as shown in Fig. \ref{Fig:model}(b).

In our simulations, we have reached linear size up to $L=128$. The inverse temperature scales as $\beta=2L$, considering the 
dynamic critical exponent $z=1$ for the two critical points studied. Typically $10^{8}$ Monte Carlo samples are taken for 
each set of parameters.

\section{ Bulk Results}
\label{sec:bulk}

\subsection{symmetry breaking phase and associated bulk phase transitions}
\begin{figure}[htb]
	\centering
	\includegraphics[width=0.46\textwidth]{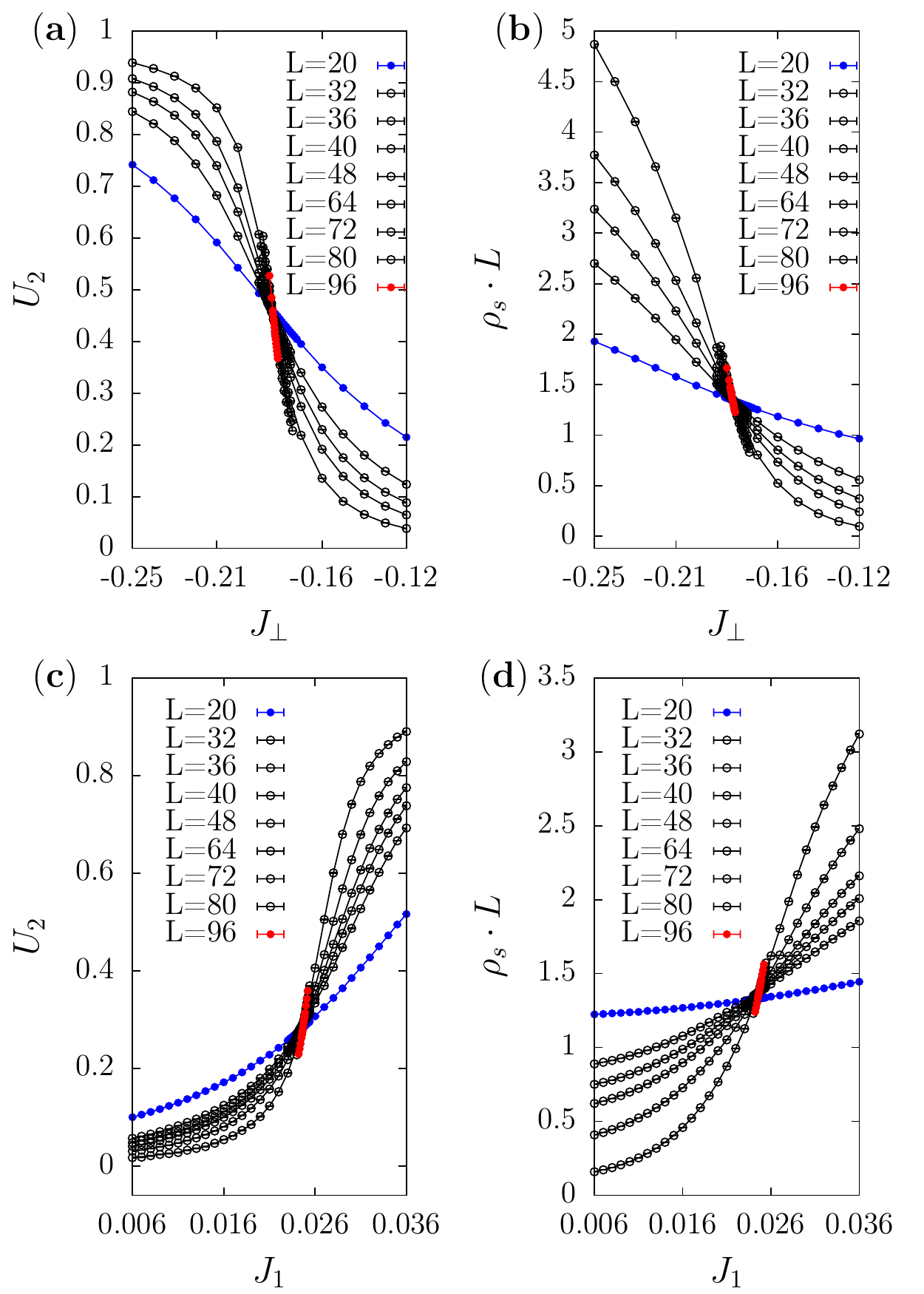}
	\caption{Binder cumulant $U_{2}$ and spin stiffness multiplied by the system size $\rho_{s}L$ versus $J_{\perp}$ or $J_{1}$
	for different system sizes. Error bars are much smaller than the symbols. (a) and (b) shows results  near the critical point $J_{\perp c}$ for setting $J_1=1$. (c) and (d) shows results near critical point $J_{1c}$ for setting $J_{\perp}=-1$.}
	\label{fig:binder}
\end{figure}

We study several physical quantities to investigate the bulk symmetry-breaking phase and related phase transitions.
In a striped magnetic phase, the spin rotational symmetry is spontaneously broken.  
The striped magnetization is used to describe this order,
\be
 m_s^z(L)=\frac{1}{L^2}\sum_i (-1)^{i_x} S_i^z,
 \ee
where $i_x= 1, 2,..., L$ is the $x$ coordinate of the spin $i$.  
The Binder cumulant $U_{2}$ \cite{Binder1981,Binder1984} is defined basing on $m_s^z$ 
\be
 U_{2}(L)= \frac{5}{6}\left(3-\frac{\langle m_{s}^{z}(L)^{4}\rangle}{\langle m_{s}^{z}(L)^{2}\rangle^{2}}\right).
 \ee
$U_{2}(L)$ converging to 1 with increasing system size indicates the existence of the striped magnetic order while tending to 
zero with increasing system size implies that the system is in the magnetically disordered phase. At the critical point, $U_2(L)$ is 
dimensionless, thus, converges to a constant.

The mean spin stiffness $\rho_s(L)$ over the $x$ and $y$ directions is calculated through the fluctuations of the winding number
of spin transporting \cite{Pollock1987, SandvikAIP}.  Similar to Binder cumulant $U_{2}$, $\rho_{s}$ is none zero if the state is magnetically 
ordered and goes to zero in the magnetically disordered phase.  The size dependence of $\rho_{s}(L)$ exactly at a QCP is expected to 
be \cite{Fisher1989}
\be
 \rho_{s}(L) \sim L^{2-(d+z)},
 \ee
with $z=1$ the dynamic exponent and $d=2$ the dimensions of the model.  
Thus, $\rho_{s}(L) L$ is expected to be dimensionless at the critical point. 

We plot $\rho_{s}(L) L$ and $U_2(L)$ as functions of $J_{\perp}$ or $J_1$ for different system sizes in Fig. \ref{fig:binder}.
Since $U_2(L)$ and $\rho_s(L) L$ are dimensionless at a QCP, the crossings of curves for different sizes roughly indicate 
the transition point.
Apparently, the model is in the striped state when $J_{\perp}$ is less than -0.18 for the setting $J_1=1$, or when $J_1$ is 
larger than 0.024 for the setting $J_{\perp}=-1$.

We adopt the standard $(L,2L)$ crossing analysis for $U_{2}(L)$ and $\rho_{s}(L) L$ curves to estimate the critical point and 
critical properties; see, e.g., the supplemental materials of \cite{shao_science}.
Let $Q$ label $U_2$ or $\rho_s L$, 
we define the finite-size estimator of the critical point $J_{c}^{(Q)}(L)$ as the crossing point of $Q$ curves for $L$ 
and $2L$, which drifts toward the critical point $J_{c}$ in the following way 
\be
J_c^{(Q)}(L)= J_c +a L^{-1/\nu-\omega}+\cdots,
\ee
where $\nu$ is the correlation length exponent, $\omega>0$ is an effective exponent for corrections to scaling, and $a$ is an unknown 
constant. 
At the crossing point $J_c^{(Q)}(L)$, the finite-size estimator of exponent $\nu$ is defined as follows 
\be
\frac{1}{\nu^{(Q)}(L)}=\frac{1}{\ln 2} \ln \left(\frac{s^{(Q)}(2L)}{s^{(Q)}(L)}\right),
\ee
where $s^{(Q)}(L)$ is the slope of the curve $Q(L)$ at $J_c^{(Q)}(L)$.
This estimator approaches the exponent $\nu$ at the speed $L^{-\omega}$;
\be
\nu^{(Q)}(L)=\nu + b L^{-\omega}+\cdots,
\ee
with $b$ an unknown constant.

The analyses basing on $U_2$ and $\rho_s L$ yield consistent estimates of $J_c$ and $\nu$
within error bars. The results with higher accuracy are selected as the final results and  
listed in Table \ref{ext1}. In particular, the final estimates of the 
two critical points are $(J_\perp=-1, J_{1c}=0.02482(2))$ and $(J_{\perp c}=-0.18271(6), J_1=1)$. 

\begin{figure}[htb]
	\centering
	\includegraphics[width=0.46\textwidth]{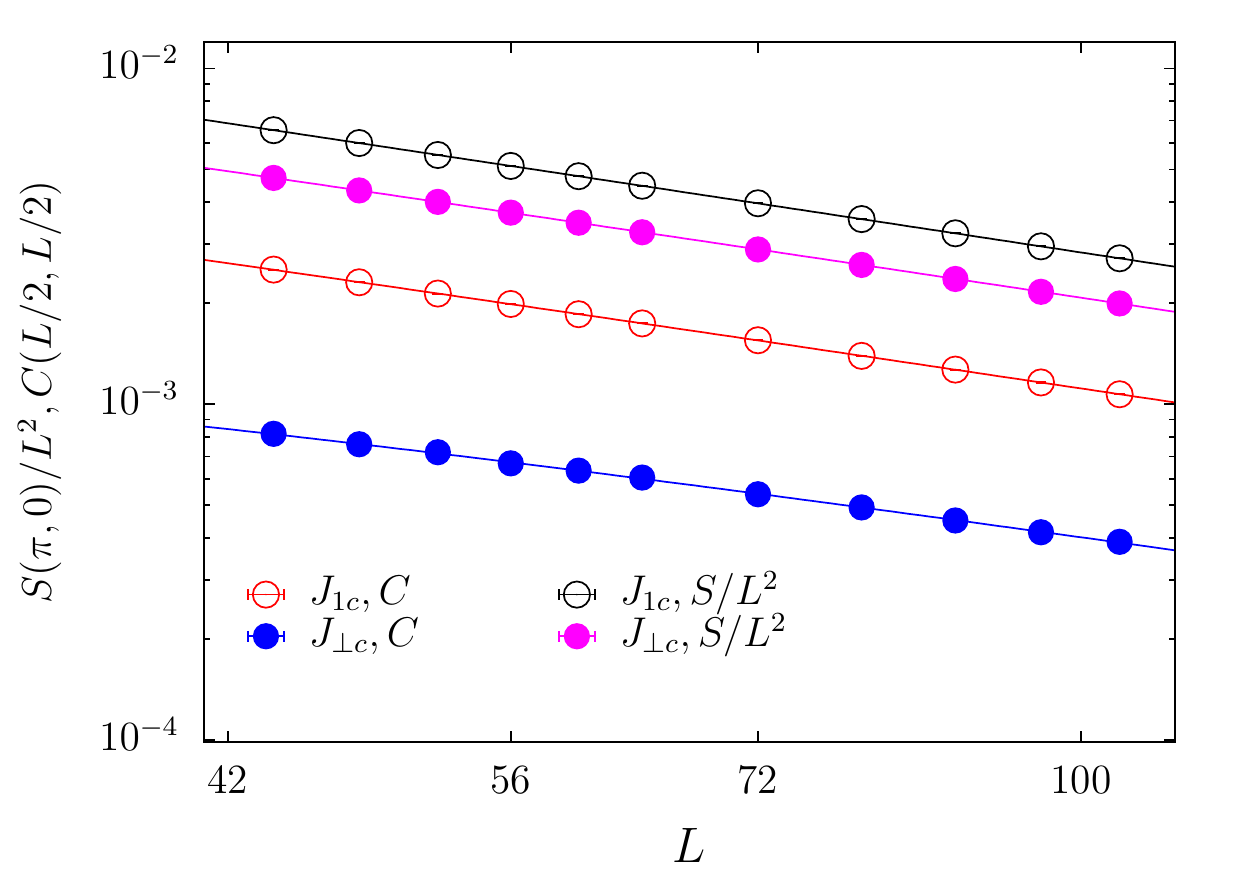}
	\caption{
	$C(L/2,L/2)$ and $S(\pi,0)/L^2$ versus system size $L$ at the quantum critical points $(-1, J_{1c})$ and $(J_{\perp c}, 1)$ on a 
	log-log scale.  Error bars are much smaller than the symbols.}
	\label{fig:bulkcandm}
\end{figure}
\begin{ruledtabular}
\begin{table*}[!t]
\caption{ Bulk critical points and exponents. Reduced $\chi^2$ (R-$\chi^2=\chi^2/{\rm d.o.f}$) and p-value of $\chi^2$ (P-$\chi^2$) are listed below the 
corresponding exponents. 	The universal exponents  obtained by field theory (FT) and by Monte Carlo simulations (MC) on the 3D classical O(3) model 
are also listed for comparison. }
\begin{tabular}{l c c c c c c  }
 	   	   &$J_{c}(U_{2})$ 	  & $J_{c}(\rho_{s}L)$ 	  & 	 $\nu (U_{2})$ & 	 $\nu (\rho_{s}L)$  	 	& $\eta$   	 &$y_{h}$  	 \\
 	   	   	
\hline
	$J_{\perp c}$			& -0.18271(6)   	&-0.185(3)  &0.72(5) &0.70(1) & 0.0350(6)   		& 2.483(2)	 \\
		R/P-$\chi^2$	    	&0.86/0.46	&0.58/0.63  &0.50/0.74 &1.09/0.36 &0.51/0.90   		& 1.5/0.12  	  \\

\hline
	$J_{1c}$			& 0.02482(2)   			&0.025(2)    & 0.73(4)   & 0.70(3)			&0.035(14)    &2.483(2)   \\

			R/P-$\chi^2$	   		&0.46/0.63	  &0.31/0.74     		&0.66/0.57	&0.36/0.88  &0.90/0.55   &1.23/0.27  \\
\hline
	FT\cite{Guida1998}			&    			&  & & 0.7073(35)   		 	&0.0355(25)    &   \\
 MC\cite{Hasenbusch2011}			&    			&  & & 0.7117(5)   		 	&0.0378(3)    &   \\

\end{tabular}
\label{ext1}
\end{table*}
\end{ruledtabular}

To further determine the universal properties of the two critical points, 
we then calculate the static spin structure factor and the spin 
correlation at the longest distance in a finite system at the two estimated critical points $(J_{\perp c}, J_1=1)$ and $( J_{\perp}=-1, J_{1c})$. 

The two quantities are defined based on the spin correlation function
\begin{equation}
 C(\mathbf{r}_{ij})=\langle S^z_i S^z_j \rangle,
\label{cij}
\end{equation}
where $\mathbf{r}_{ij}$ is the vector from site $i$ to $j$. 
The static spin structure factor at wave vector $(\pi, 0)$ is used to describe the striped order, which is defined as follows
 \begin{equation}
 S(\pi,0)=\sum_{\mathbf{r}_{ij}} (-1)^{i_x-j_x} C(\mathbf{r}_{ij}),
\label{spipi}
\end{equation}
 where $i_x,j_x=1,2,...,L$ are the $x$ coordinate of the spin $i$ and $j$, respectively.
The spin correlation function $C(L/2,L/2)$ averages $C(\mathbf{r}_{ij})$ 
between two spins $i$ and $j$ at the longest distance $\mathbf{r}_{ij}=(L/2,L/2)$.

Using $S(\pi,0)$  and $C(L/2,L/2)$, we  
extract the scaling dimension $y_{h}$ of the striped magnetic field $h$ conjugating to the striped magnetization $m_s^z$ and the anomalous 
dimension  $\eta$. 

At a QCP, $S(\pi,0)$  and $C(L/2,L/2)$ satisfy the following finite-size scaling forms
\begin{equation}
 S(\pi,0)/L^2\sim L^{-2(d+z-y_{h})}(1+b_1 L^{-\omega_1}),
\label{spi}
\end{equation}
and
\begin{equation}
 C(L/2,L/2)\sim L^{-(d+z-2+\eta)}(1+b_2 L^{-\omega_2}),
\label{cb}
\end{equation}
respectively, in which $d=2$ is the spatial dimension, $z=1$ is the dynamical critical exponent.
$\omega_1$ and $\omega_2$ are the effective exponents for corrections to scaling.
The two exponents $y_h$ and $\eta$ are not independent and are expected to obey the following scaling relation
\begin{equation}
\eta=d+z+2-2 y_{h}.
\label{scalingb}
\end{equation}

Figure \ref{fig:bulkcandm} shows the numerical results of $ S(\pi,0)/L^2$ and $C(L/2, L/2)$ as functions of system size $L$ at two critical points.
Fitting  Eqs. (\ref{spi}) and (\ref{cb}) to the data of $S(\pi,0)/L^2$ and $C(L/2,L/2)$, respectively, we obtain the critical
exponents $y_{h}$ and $\eta$ at the two critical points, respectively. 
In these fits, we find that setting $\omega_1=1$ and $\omega_2=1$ yield good fits. 
The results  are also presented in Table \ref{ext1}. 
The two pairs of $y_{h}$ and $\eta$ agree well  and satisfy the scaling relations Eq. (\ref{scalingb}) within error bars.

Comparing with the best known exponents of the 3D O(3) universality class \cite{Guida1998,Hasenbusch2011}, we conclude that both 
critical points belong to the (2+1)-D O(3) universality class.  
Combined with the previous numerical results \cite{Matsumoto2001, Wang2022}, these results further support that the topological order does not change 
the universality class of the bulk phase transition, described by the Landau symmetry-breaking paradigm.

\subsection{The Haldane phases and their surface states}
\label{subsec:Haldane}

	\begin{figure}[htb]
	\centering
	\includegraphics[width=0.46\textwidth]{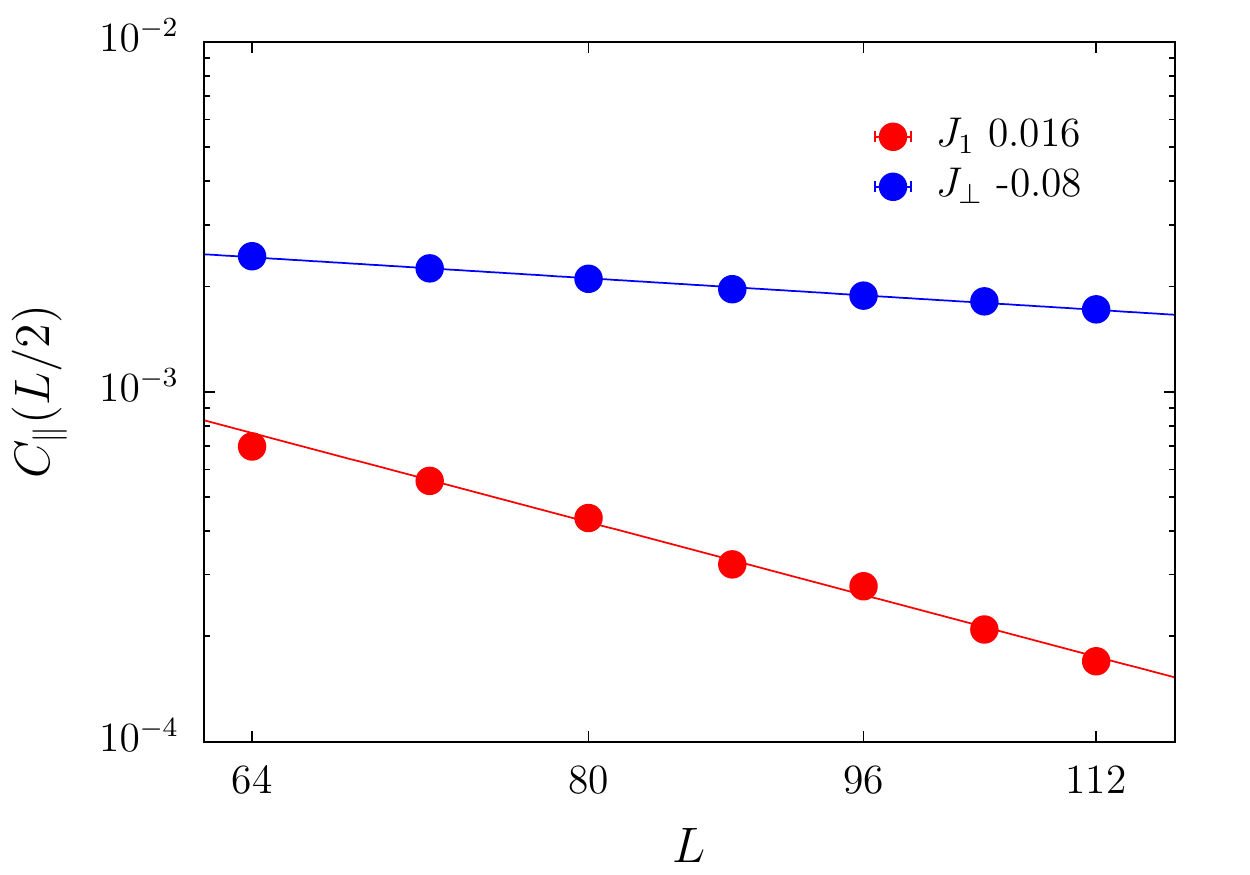}
	\caption{
	Surface correlation $C_{\parallel}(L/2)$ versus system size $L$ on a log-log scale.
$C_{\parallel}(L/2)$ are calculated at $(J_{\perp}=-0.08, J_1=1)$ in the DHFM phase and $(J_{\perp}=-1,J_1=0.016)$ in the UH phase, respectively. 
 Algebraically decaying with $L$ is observed for both cases, showing two gapless surface states. }
	\label{Fig:surfacegapbulk}
\end{figure}

We now look into the properties of the nonmagnetic phase closer. 

An $S=1/2$ diagonal ladder, shown inside the dashed rectangular box in Fig. \ref{Fig:model}(a), is the composite spin representation of an $S=1$ 
chain\cite{diagonal2000}. Its ground state is described by the AKLT state \cite{AKLT}, a typical configuration of which is illustrated in
Fig. \ref{Fig:diagram}(a) at small $|J_\perp|$.
An usual ladder with FM rung coupling $J_\perp<0$, shown inside the dashed rectangular box in Fig. \ref{Fig:model}(b), also 
behaves like an $S=1$ chain with a gap to excitations. Therefore, the ground state can also be represented by the AKLT state, with a typical valence
bond configuration shown in Fig. \ref{Fig:diagram}(b) at small $J_1$.
These two gapped states, corresponding to odd integer spin chains, are well-known SPT Haldane phases \cite{Haldane, Affleck} with 
the symmetry-protected topological order\cite{XiaoGang2009, Pollmann2010}. With open boundaries, the ground states have two spin-1/2 spins localized
at the ends of the ladder. 

When the usual ladders with FM rung couplings are weakly coupled, the system is still in the SPT Haldane phase (UH) due to 
the gap, as illustrated in Fig. \ref{Fig:diagram}(b). Similarly,  weakly coupled diagonal ladders stay in the SPT Haldane phase due to the 
gap. The antiferromagnetically coupled
case (DHAF), shown in Fig. \ref{Fig:diagram}(a)), has been studied in \cite{Wang2022}. In the current paper, we focus on the FM coupled 
case (DHFM), also illustrated in Fig.\ref{Fig:diagram} (a). 

In 1D, Haldane phases are often characterized by a “hidden” nonlocal order parameter, the so-called string order \cite{Nijs1989, diagonal2000}. 
However, the string order parameter is fragile to arbitrary weak higher-dimensional couplings between such chains or ladders 
\cite{Frank2010, Anfuso2007}, 
which has been verified numerically recently in the 2D coupled spin-1 Haldane chains (CHCs) model\cite{Zhu2021} and in the 2D spin-1/2 CDLs 
model \cite{Wang2022}. 
Nevertheless, the hallmark of the SPT phase is the presence of nontrivial surface states that are gapless or degenerate \cite{wen2017}, which 
have been shown to be present in the 2D CHCs \cite{Zhu2021} and in the 2D CDLs\cite{Wang2022}. We will show below that this is true also 
for the weakly FM-coupled diagonal ladders and weakly diagonally coupled  usual ladders with FM rung couplings. 

To study the surface states on the surfaces perpendicular to the ladders (see Fig. \ref{Fig:model} (b)), we calculate the surface parallel 
correlation $C_\parallel(L/2)$, which averages 
$C(\mathbf{r}_{ij})$ between two surface spins $i$ and $j$ at the longest distance $L/2$. 

Figure \ref{Fig:surfacegapbulk} shows $C_\parallel(L/2)$ at $(J_\perp=-1, J_1=0.016)$ sitting in the UH phase and that at $(J_\perp=-0.08, J_1=1)$ 
sitting in the DHFM phase.
In both cases, we see that $C_\parallel(L/2)$ decays with system size $L$ in a power law, 
\begin{equation}
    C_\parallel(L/2) \sim  L^{-p}.
    \label{gapless_surface}
\end{equation}
We find $p=2.6(1)$ for the former case and  
$p=0.62(2)$ for the latter case, 
meaning that both surface states are gapless. 

Similar power-law decaying was found on the surface of the DHAF phase \cite{Wang2022}. 
Notably, the surface configurations are different there, as already mentioned in the introduction:  
In the DHFM phase, the inter-ladder couplings are ferromagnetic $J_{\perp}<0$ instead of antiferromagnetic in the DHAF phase, while in the UH 
phase, the ladders are coupled by diagonal AF couplings, inducing effective FM couplings between two neighboring spins in two ladders.  
The gapless surface states demonstrate that the naive idea of thinking of the surface as a ferromagnetic chain of $S=1/2$ spins localized  at the ends 
of the ladders, which is gapped,  is incorrect. 
The gapless surface states purely reflect the topological order of the bulk.  

\section{ Surface critical behaviors}
\label{sec:scb}

We now study the surface critical behaviors  at the two bulk critical points at $(J_\perp=-1, J_{1c})$ and $(J_{\perp c}, J_1=1)$, respectively.

Besides the surface correlation $C_\parallel(L/2)$, we calculate another spin correlation $C_\perp(L/2)$ and 
the  squared ferromagnetic surface magnetization  $m^2_{1}(L)$. 
$C_\perp(L/2)$ averages $C(\mathbf{r}_{ij})$ between spin $i$ fixed on the surface and spin $j$
located at the center of the bulk, with $\mathbf{r}_{ij}$ perpendicular to the surface and $|j-i|=L/2$.
The surface magnetization $m_1(L)$ is defined as follows
\be
m_{1}(L) =\frac{1}{L}\sum_{i \in {\rm surface}}  S_i^z,
\ee
where the summation is restricted for spins on the surface.

At a bulk critical point, for the ordinary, special, and extraordinary SCBs, the finite-size scaling behavior of the two correlations is 
characterized by two 
anomalous dimensions $\eta_\parallel$ and $\eta_\perp$, respectively\cite{Binder1974};
\begin{equation}
C_{\parallel}(L/2) =C_\parallel+a_1 L^{-(d+z-2+\eta_{\parallel})}+a_2 L^{-1}+\cdots,
\label{cs1}
\end{equation}
and
\begin{equation}
C_{\perp}(L/2) = b_1 L^{-(d+z-2+\eta_{\perp})}+b_2 L^{-1}+\cdots,
\label{cs2}
\end{equation}
where  $a_i$ and $b_i$ are unknown constants, $1/L$ terms are the leading correction to scaling due to analytic contributions.
$C_\parallel=0$  in ordinary or special SCBs; $C_\parallel \neq 0$ characterizes the long-range order on the surface in an
extraordinary SCB. 
The squared surface magnetization follows  the scaling form\cite{Binder1974}
\begin{equation}
m^2_{1}(L) = m^2_1+ c_1 L^{-2(d+z-1-y_{h1})} +c_2 L^{-1} +\cdots,
\label{m1}
\end{equation}
where  $y_{h1}$ is the scaling dimension of the surface field $h_1$,  $1/L$ is the leading contribution from
analytic terms,  and $c_i$  are unknown constants. 
$m^2_1$ should be equal to $C_\parallel$, corresponding to the squared magnetization on the
surface in an extraordinary SCB. 
For our model, $d=2$ and $z=1$.

The three exponents $y_{h1}$, $\eta_{\parallel}$, and $\eta_{\perp}$ are different for ordinary, special, and extraordinary transitions; but are related through the following scaling relations:\cite{Diehl}
 \begin{equation}
2\eta_\perp = \eta_\parallel +\eta  
\label{scalings1} 
\end{equation}
and
\begin{equation}
\eta_\parallel = d+z - 2y_{h1},
\label{scalings2}
\end{equation} 
with $\eta$ the anomalous magnetic scaling dimension of the bulk critical point in the $d+z$ spacetime. We use these physical quantities in the remainder of this section to examine SCBs. Two extraordinary  SCBs at different bulk critical are found. 

For the 3D model with a continuous symmetry-breaking critical point, classical theory does not support extraordinary SCB since its 2D surface 
can not order; hence, no special SCB is available. 
However, when quantum mechanics sets in, it is possible that the (1+1)-D surface becomes long-range ordered due to coupling to a (2+1)-D O(3) QCP \cite{JianCM2021} and exhibits extraordinary SCBs.  Numerical results have found such extraordinary SCBs in different models \cite{Ding2018, Ding2021}.

A novel extraordinary-log SCB was proposed for a surface critical state in 3D O($N$) critical point by Metlitski \cite{Metlitski2020} for $2\leq N < 
N_c$, in which the spin correlation decay logarithmically 
\begin{equation}
    C_\parallel(L) \propto \log(L/L_0)^{-q},
    \label{cparalog}
\end{equation}
where $L_0$ is a nonuniversal constant. The surface magnetization $m^2_1(L)$ also decays logarithmically with the same exponent $q$.
This extraordinary-log SCB has been verified numerically in the classical 3D O(3) model \cite{ParisenToldin2021} and 3D O(2) model \cite{Hu2021}.
This behavior was also found on the dangling-chain surface in spin-1 dimerized Heisenberg model\cite{Zhu2021b}, suggesting that such SCB may 
apply to 2D QCP for the integer spin model.

Previous studies \cite{Zhu2021, Wang2022} on QCPs between SPT phases and O(3) symmetry-breaking phase in (2+1)-D have found that the gapless 
edge modes of the SPT phases 
lead to nonordinary SCBs unexpected according to the quantum-classical correspondence. Such nonordinary SCBs show no surface order and have 
different exponents from the ordinary SCB of the 3D O(3) UC. The exponents are 
similar  to the exponents of the SCBs
found at the (2+1)-D O(3) QCPs separating the topological trivial product states and the symmetry-breaking phase\cite{Ding2018, Weber2018, Ding2021, 
Wang2022}, where the surface is a 
dangling $S=1/2$ chain at the product state, thus, is also gapless according to the Lieb-Schultz-Mattis theorem. 

In this work, we have shown that the QCPs between the striped phase and the two SPT phases (the UH and the DHFM phases) belong to the 3D O(3) UC. 
Both of them have gapless edge modes on the surfaces formed by the ends of the ladders, which are not dangling. 
In the next two subsections, we will check the various SCBs listed above on these surfaces
and show that the gapless edge modes lead to unexpected extraordinary SCBs, which suggests that the two SPT states, i.e., DHFM and UH,
are topologically different from the SPT state (DHAF) studied in \cite{Wang2022} and the SPT state of AF coupled spin-1 chain\cite{Zhu2021}.

\begin{figure}[htb]
	\centering
	\includegraphics[width=0.46\textwidth]{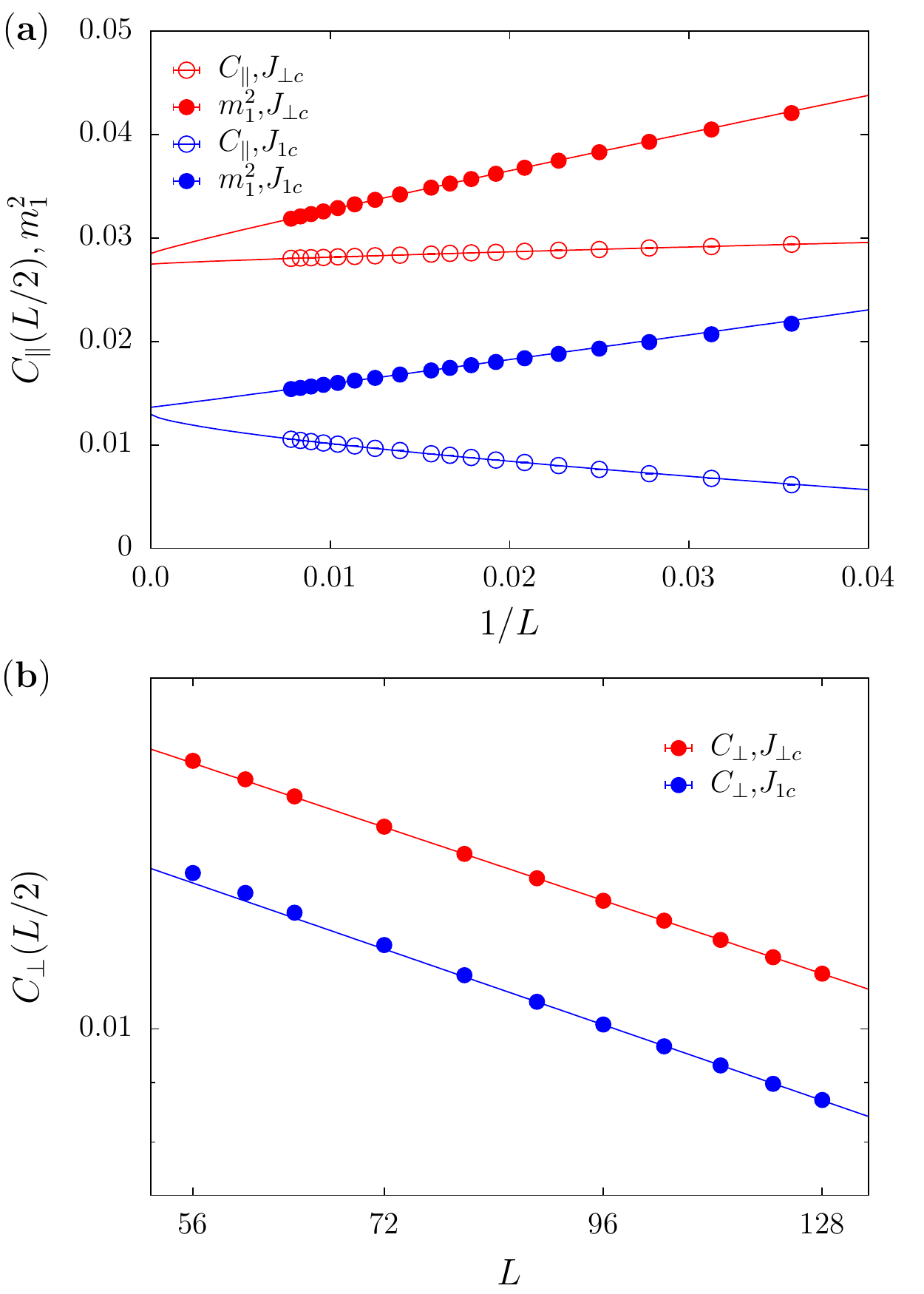}
	\caption{(a) Squared surface ferromagnetic magnetization $m_1^2$  and the correlations $C_{\parallel}(L/2)$ versus $1/L$ and
	(b) The correlations  $C_{\perp}(L/2)$ versus system size $L$ on a log-log scale  at two bulk critical points. 
	Lines are fits according to Eq. (\ref{cs1})-(\ref{m1}) }
	\label{fig:surface}
\end{figure}

\begin{ruledtabular}
\begin{table}[!t]
\caption{ Finite-size scaling analysis of $C_{\parallel}(L/2)$, $m^2_{1}(L)$ and $C_{\perp}(L/2)$ at bulk critical point $J_{1c}$ to obtain surface critical exponents $\eta_{\parallel}$, $y_{h1}$ and $\eta_{\perp}$, respectively. Correction term $L^{-1}$ included in the fitting formula Eq.(\ref{cs1}, (\ref{cs2}), and (\ref{m1}) are denoted by "Yes", not included are denoted by "No". Reduced $\chi^2$ (R-$\chi^2$) and p-value of $\chi^2$ (P-$\chi^2$) are also listed. 	 }
\begin{tabular}{l c c c c  }
 	   	   &$L_{\min}$ 	  & $\eta_{\parallel}$ 	  & R/P-$\chi^2$	&b 	 \\
 	   	   	
\hline
	Yes			& 56   	&-0.33(5)  &0.98/0.45  &-0.05(1)	 \\
					& 64   	&-0.34(8)  &1.15/0.32  &-0.05(2)   \\
          		& 72   	&-0.3(1)  &1.38/0.22  & -0.05(4) \\

\hline
        No			& 72   	&-0.29(1)  &1.18/0.30 & -0.07(1)	 \\
					& 88   	&-0.30(2)  &0.76/0.55 &  -0.06(1)	 \\
          		& 96   	&-0.30(3)  &1.0/0.39 &  -0.06(1)	 \\
\hline
\hline
	&$L_{\min}$ 	  & $y_{h1}$ 	  & R/P-$\chi^2$	 	 \\
 \hline
        Yes			& 48  	&1.69(8)  &1.34/0.20 &0.04(1)  \\
					& 56   	&1.7(1)  &1.08/0.37 &0.036(6) \\
          		& 64   	&1.7(2)  &0.42/0.86 &0.03(1) \\
     \hline
        No			& 88   	&1.641(4)  &1.71/0.14 	&	0.09(1) \\
					& 96   	&1.64(1)  &1.50/0.21 &	0.08(1)	 \\
          		& 104  	&1.65(1)  &0.53/0.59 &	0.08(1)	 \\
        
\hline
\hline
	&$L_{\min}$ 	  & $\eta_{\perp}$ 	  & R/P-$\chi^2$	 	 \\
 \hline
        Yes			& 80  	&-0.56(4)  &0.55/0.70 &0.06(2)	 \\
					& 88   	&-0.55(6)  &0.61/0.61 & 0.07(2)	 \\
          		& 96   	&-0.6(2)  &0.49/0.61  &	0.05(2) \\
     \hline
        No			& 80   	&-0.48(1)  &1.88/0.09 &0.110(2)	 \\
					& 88   	&-0.48(1)  &0.83/0.51 &0.108(2)	 \\
          		& 96  	&-0.48(1)  &1.00/0.39 &0.11(1)	 \\
\end{tabular}
\label{ext2}
\end{table}
\end{ruledtabular}
\subsection{Surface critical behaviors at the QCP between the UH and the striped phases}

We first study the surface critical behaviors associated with the bulk critical point $(J_\perp=-1, J_{1c})$ separating the  striped  magnetic 
ordered phase from the SPT UH phase. 

The numerical results of $C_{\parallel}(L/2)$ and $m^2_{1}(L)$ as functions of size $L$ are graphed in 
Fig. \ref{fig:surface}(a), $C_{\perp}(L/2)$ as a function of $L$ are plotted in Fig. \ref{fig:surface}(b).

Evidently, the extraordinary-log SCBs are excluded since $C_\parallel(L/2)$ increases with size $L$.
We then try to analysis $C_\parallel(L/2)$ according to Eq. (\ref{cs1}). 

Suppose $\eta_\parallel<0$, we can ignore the $L^{-1}$ term for large $L$. Or, on the other hand, if $\eta_\parallel>0$, the second term
in Eq. (\ref{cs1}) can be ignored for large $L$. In either case, as $C_\parallel(L/2)$ increases with size $L$, it is evident 
that $C_\parallel >0$. 
Fitting Eq. (\ref{cs1}) with $\eta_{\parallel}$, $C_\parallel$, and $a_1$ free, ignoring the $L^{-1}$ term, we obtain   $C_\parallel=0.0124(6)$.
However, $\eta_{\parallel}$ is found close to 0, making it difficult to separate the singular parts from the analytic term $1/L$. 
We have also tried to include the $1/L$ term in Eq. (\ref{cs1}) in the fitting. This does not lead to a meaningful estimate of $\eta_\parallel$.

We fit $m^2_1(L)$ according to Eq. (\ref{m1}) with $y_{h1}$, $m^2_1$, and $c_1$ free,  ignoring analytic terms.    
We find $m^2_1=0.013(1)$, in good agreement  with $C_\parallel$ within error bars, supporting long-range order on the 
surface.  Unfortunately, in this fitting, we find $y_{h1}$ very close to 1.5. Again, it is difficult to separate the singular parts from the 
analytic correction $1/L$.

With the long-range order on the surface determined,  
it is tempting to fix $C_\parallel=m^2_1=0.0127$ in the fitting formula Eq. (\ref{cs1}) for $C_\parallel(L)$ and Eq. (\ref{m1}) for 
$m^2_1(L)$, respectively, 
and fit for $\eta_\parallel$ and $y_{h1}$ again. By gradually excluding small sizes, we achieve stable fits for both $C_\parallel(L)$ and 
$m^2_1(L)$, with the analytic correction $1/L$ term included and not included, respectively.  
The details of the fitting procedure are presented in Table \ref{ext2}. Our final estimate of the exponent 
$\eta_{\parallel}$ is $\eta_{\parallel}=-0.30(3)$ and the exponent $y_{h1}$ is $y_{h1}=1.65(1)$. The two exponents obey the scaling 
relation Eq.~(\ref{scalings2}). 
However, these results are based on assuming values of $C_\parallel$ and $m^2_1$; one should be cautious about the reliability of these estimates.

The finite-size scaling form Eq. (\ref{cs2}) is used to fit the data of $C_{\perp}(L/2)$. The stable fits are obtained for sufficient large sizes, 
and are also listed in Table \ref{ext2}. 
Our final estimate of the exponent $\eta_{\perp}$ is $\eta_{\perp}=-0.48(1)$. The result is consistent 
with $\eta_{\perp}=-0.5050 (10)$ found in the extraordinary transition on the dangling ladder surface of the 2D staggered Heisenberg 
model \cite{Ding2018}. 

Noting that the exponents $\eta_{\parallel}=-0.30(3)$ and  $\eta_{\perp}=-0.48(1)$ violate the relation in Eq. (\ref{scalings1}).
This fact makes the value of $\eta_\parallel=-0.30(3)$, as well as $y_{h1} = 1.65(1)$, more doubtful.

\subsection{Surface critical behaviors at the QCP between the DHFM and the striped phases}

We then check the SCBs at the critical point $(J_{\perp c}, J_1=1)$ where the SPT DHFM phase transfers to the striped  magnetic ordered phase.

The numerical results of  $C_\parallel(L/2)$ and $m^2_{1}$  versus $L$ are plotted in Fig. \ref{fig:surface}(a). 

The SPT DHFM state is the ground state of weakly FM-coupled diagonal ladders.
The diagonal ladder, like the usual ladder with FM rungs, behaves like a spin-1 chain,
and its VB ground state is also odd \cite{diagonal2000}. 
As discussed in the introduction, both the DHFM state and the UH state can be regarded as FM-coupled spin-1 chains. 
Therefore, we expect qualitatively the same SCBs as those at the UH and the striped phase QCP  and exclude the extraordinary-log scenario.
Nevertheless, we have checked the extraordinary-log behaviors. 
Fitting $C_\parallel(L/2)$ according to Eq. (\ref{cparalog}) yields $q=0.09(20)$, similar fitting to $m_1^2(L)$ yields $q=0.156(7)$. Hence, although we
can not completely exclude the extraordinary-log SCB,  we conclude that such behavior is unlikely. 

Fitting these
data according to Eqs. (\ref{cs1}) and (\ref{m1}) with the same procedure as at critical point $(J_\perp=-1, J_{1c})$, we obtain  $C_\parallel=0.0275(3)$ and $m^2_1=0.028(2)$, which are consistent within error bars. 
This indicates the existence of a long-range order on the surface. Again, $\eta_\parallel$ is found close to 0 and $y_{h1}\approx 1.5$. It is
difficult to separate the singular parts from the analytic correction. 

\begin{ruledtabular}
\begin{table}[!t]
\caption{ Finite-size scaling analysis of $C_{\parallel}(L/2)$, $m^2_{1}(L)$, and $C_{\perp}(L/2)$ at bulk critical point $J_{\perp c}$ to obtain surface critical exponents $\eta_{\parallel}$, $y_{h1}$, and $\eta_{\perp}$, respectively. Fits with correction terms $L^{-1}$ included
are denoted by "Yes", those without $L^{-1}$ term are denoted by "No". 
Reduced $\chi^2$ (R-$\chi^2$) and p-value of $\chi^2$ (P-$\chi^2$) are also listed. 	 }
\begin{tabular}{l c c c c  }
 	   	   &$L_{\min}$ 	  & $\eta_{\parallel}$ 	  & R/P-$\chi^2$	&b 	 \\
 	   	   	
\hline
	Yes			& 44   	&-0.26(13)  &0.34/0.97  &0.018(9) 	 \\
				& 48   	&-0.28(16)  &0.37/0.96 &0.02(1)	 \\
          		& 56   	&-0.5(4)  &0.26/0.97 &0.01(1)	 \\

\hline
        No		& 72   	&-0.20(4)  &0.23/0.96 	& 0.026(3) \\
				& 80   	&-0.19(5)  &0.27/0.93  & 0.027(4)	 \\
          		& 88   	&-0.22(7)  &0.27/0.90 &0.02(1)	 \\
\hline
\hline
	&$L_{\min}$ 	  & $y_{h1}$ 	  & R/P-$\chi^2$	 	 \\
 \hline
        Yes		& 72  	&1.70(5)  &0.32/0.89	&0.07(2) \\
				& 80   	&1.71(6)  &0.39/0.81 	&0.06(2) \\
          		& 88   	&1.7(1)  &0.52/0.66 	 &0.06(4)\\
     \hline
        No		& 72   	&1.63(1)  &1.66/0.12 	&0.16(1) \\
				& 80   	&1.63(1)  &1.03/0.39 &0.15(1)	 \\
          		& 96  	&1.64(1)  &0.70/0.55 &0.15(1)	 \\
 \hline
\hline
	&$L_{\min}$ 	  & $\eta_{\perp}$ 	  & R/P-$\chi^2$	 	 \\
 \hline
        Yes		& 80  	&-0.52(2)  &0.25/0.91	 &0.114(4)\\
				& 88   	&-0.52(3)  &0.34/0.80 	& 0.113(8)\\
          		& 96   	&-0.53(4)  &0.45/0.64 	 &0.11(2)\\
     \hline
        No		& 80   	&-0.497(1)  &0.86/0.51 &0.128(1)	 \\
				& 88   	&-0.498(2)  &0.52/0.72 &0.127(1)	 \\
          		& 96  	&-0.499(2)  &0.52/0.67 	&0.127(1) \\

\end{tabular}
\label{ext3}
\end{table}
\end{ruledtabular}

 We then fix $C_\parallel=m^2_1=0.0275$  in the fitting of Eq. (\ref{cs1}) and (\ref{m1}) to $C_\parallel(L/2)$ and $m^2_1(L)$, respectively. 
 We achieve stable fit with $1/L$ term included or not included in the fitting procedure for sufficiently large $L_{min}$. The results are listed 
 in Table \ref{ext3}.  Our final estimate of the exponents are $\eta_{\parallel}=-0.22(7)$ and $y_{h1}=1.64(1)$. Two  exponents  satisfy the
scaling relation Eq. (\ref{scalings2}) and are consistent with the results at $(J_\perp=-1, J_{1c})$.
Again, we have reservations about the values of these exponents.


The numerical results of  $C_{\perp}(L/2)$  as functions of $L$ are plotted in Fig. \ref{fig:surface}(b). Stable fits are obtained by gradually
excluding data of sizes smaller than $L_{min}$. The procedure is presented in 
Table \ref{ext3}. Our final estimate is $\eta_{\perp}=-0.498(2)$. The exponent  is consistent 
with the results at $J_{1c}$  as well as the  $\eta_{\perp}=-0.5050 (10)$ found in the extraordinary transition on a special surface of the 
2D staggered Heisenberg model \cite{Ding2018}.  

Apparently, $\eta_{\parallel}$ and  $\eta_{\perp}$ also challenge the relation in Eq. (\ref{scalings1}). 
This fact makes the value of $\eta_\parallel=-0.22(7)$, as well as $y_{h1} = 1.64(1)$, suspectable, however,  they agree with the corresponding
exponents found at the critical point $(J_\perp=-1, J_{1c})$.

\section{Discussion and Conclusion}
\label{sec:conclsn}
Using quantum Monte Carlo simulations, we have studied the spin-1/2 Heisenberg model on the 2D CDLs lattice  with tunable ferromagnetic 
inter-ladder couplings. The model can also be considered the spin-1/2 Heisenberg model on the 2D coupled FM usual ladders with tunable 
antiferromagnetic diagonal couplings. We have studied the phases and phase transitions when the inter-ladder
coupling is tuned. We have shown that the model realizes two 2D SPT Haldane phases, the UH and the DHFM phases, when the two kinds of ladders 
are weakly coupled, respectively, and that the model enters the striped magnetic ordered phase when the couplings of ladders are strong enough. 
We have demonstrated that the two quantum critical points separating the SPT phases and the striped phase are in the 3D O(3) universality class; 
the topological properties of the SPT phases do not  affect the universal properties of the bulk phase transitions.

We have also studied the surface states of the two SPT phases and found gapless surface modes on the surfaces formed by the ends of the ladders. 
Unlike the surface configurations of the coupled Haldane chains\cite{Zhu2021} and that of the 
spin-1/2 CDLs model with ladder coupling $J_{\perp}>0$ \cite{Wang2022} in the corresponding SPT Haldane phases, where the spin-1/2 excitations at 
the ends of the chains/ladders are coupled  antiferromagnetically, here the spin-1/2 excitations at the ends of the ladders 
are coupled ferromagnetically. However, these surface 
excitations should not be viewed as a 1D ferromagnetic $S=1/2$ chain, which is gapped. Instead, the gapless surface modes in both SPT 
phases should be considered to originate purely from the bulk-edge correspondence of the topologically ordered bulk state.

We have focused on the SCBs at the two bulk critical points. 
Unlike in the CHCs model and the AF-coupled spin-1/2 CDLs model, where the gapless edge modes of the SPT phases induce nonordinary SCBs, we 
found that the gapless surface states of the UH and the DHFM SPT states of the current model lead to 
extraordinary SCBs at the two bulk QCPs with magnetically ordered surfaces.
This finding reveals that the SPT state of the FM-coupled CDLs model and the SPT state of the diagonally AF-coupled FM usual ladders are 
topologically different from the SPT state of the AF-coupled CDLs model
according to the bulk-edge correspondence of a topologically ordered state.

Ultimately, we would like to mention that an SPT state leading to extraordinary SCB at a QCP is a surprise. To our knowledge, there is 
no theoretical explanation for this. Therefore, further investigations are called for.

\begin{acknowledgments}
We thank Prof. H.Q. Lin and Dr. Wenjing Zhu for their valuable discussions.
This work was supported by the National Natural Science Foundation of China under Grant No.~12175015 and No.~11734002. 
The authors acknowledge the support the Super Computing Center of Beijing Normal University extended.
\end{acknowledgments}

\bibliography{ref.bib}

\begin{thebibliography}{47}%
\makeatletter
\providecommand \@ifxundefined [1]{%
 \@ifx{#1\undefined}
}%
\providecommand \@ifnum [1]{%
 \ifnum #1\expandafter \@firstoftwo
 \else \expandafter \@secondoftwo
 \fi
}%
\providecommand \@ifx [1]{%
 \ifx #1\expandafter \@firstoftwo
 \else \expandafter \@secondoftwo
 \fi
}%
\providecommand \natexlab [1]{#1}%
\providecommand \enquote  [1]{``#1''}%
\providecommand \bibnamefont  [1]{#1}%
\providecommand \bibfnamefont [1]{#1}%
\providecommand \citenamefont [1]{#1}%
\providecommand \href@noop [0]{\@secondoftwo}%
\providecommand \href [0]{\begingroup \@sanitize@url \@href}%
\providecommand \@href[1]{\@@startlink{#1}\@@href}%
\providecommand \@@href[1]{\endgroup#1\@@endlink}%
\providecommand \@sanitize@url [0]{\catcode `\\12\catcode `\$12\catcode
  `\&12\catcode `\#12\catcode `\^12\catcode `\_12\catcode `\%12\relax}%
\providecommand \@@startlink[1]{}%
\providecommand \@@endlink[0]{}%
\providecommand \url  [0]{\begingroup\@sanitize@url \@url }%
\providecommand \@url [1]{\endgroup\@href {#1}{\urlprefix }}%
\providecommand \urlprefix  [0]{URL }%
\providecommand \Eprint [0]{\href }%
\providecommand \doibase [0]{https://doi.org/}%
\providecommand \selectlanguage [0]{\@gobble}%
\providecommand \bibinfo  [0]{\@secondoftwo}%
\providecommand \bibfield  [0]{\@secondoftwo}%
\providecommand \translation [1]{[#1]}%
\providecommand \BibitemOpen [0]{}%
\providecommand \bibitemStop [0]{}%
\providecommand \bibitemNoStop [0]{.\EOS\space}%
\providecommand \EOS [0]{\spacefactor3000\relax}%
\providecommand \BibitemShut  [1]{\csname bibitem#1\endcsname}%
\let\auto@bib@innerbib\@empty
\bibitem [{\citenamefont {Landau}\ and\ \citenamefont
  {Lifshitz}(1980)}]{landau1980statistical}%
  \BibitemOpen
  \bibfield  {author} {\bibinfo {author} {\bibfnamefont {L.}~\bibnamefont
  {Landau}}\ and\ \bibinfo {author} {\bibfnamefont {E.}~\bibnamefont
  {Lifshitz}},\ }\href {https://books.google.com/books?id=dEVtKQEACAAJ} {\emph
  {\bibinfo {title} {Statistical Physics}}},\ \bibinfo {number} {5}\ (\bibinfo
  {publisher} {Elsevier Science},\ \bibinfo {year} {1980})\BibitemShut
  {NoStop}%
\bibitem [{\citenamefont {Wilson}\ and\ \citenamefont
  {Kogut}(1974)}]{WILSON197475}%
  \BibitemOpen
  \bibfield  {author} {\bibinfo {author} {\bibfnamefont {K.~G.}\ \bibnamefont
  {Wilson}}\ and\ \bibinfo {author} {\bibfnamefont {J.}~\bibnamefont {Kogut}},\
  }\bibfield  {title} {\bibinfo {title} {The renormalization group and the
  $\epsilon$ expansion},\ }\href
  {https://doi.org/https://doi.org/10.1016/0370-1573(74)90023-4} {\bibfield
  {journal} {\bibinfo  {journal} {Physics Reports}\ }\textbf {\bibinfo {volume}
  {12}},\ \bibinfo {pages} {75} (\bibinfo {year} {1974})}\BibitemShut {NoStop}%
\bibitem [{\citenamefont {Cardy}(1996)}]{Cardy}%
  \BibitemOpen
  \bibfield  {author} {\bibinfo {author} {\bibfnamefont {J.}~\bibnamefont
  {Cardy}},\ }\href {https://doi.org/10.1017/CBO9781316036440} {\emph {\bibinfo
  {title} {{S}caling and renormalization in statistical physics}}},\ \bibinfo
  {series} {Cambridge lecture notes in physics}, Vol.~\bibinfo {volume} {5}\
  (\bibinfo  {publisher} {Cambridge University Press},\ \bibinfo {address}
  {Cambridge},\ \bibinfo {year} {1996})\ p.\ \bibinfo {pages} {238
  pages}\BibitemShut {NoStop}%
\bibitem [{\citenamefont {Binder}\ and\ \citenamefont
  {Hohenberg}(1974)}]{Binder1974}%
  \BibitemOpen
  \bibfield  {author} {\bibinfo {author} {\bibfnamefont {K.}~\bibnamefont
  {Binder}}\ and\ \bibinfo {author} {\bibfnamefont {P.~C.}\ \bibnamefont
  {Hohenberg}},\ }\bibfield  {title} {\bibinfo {title} {Surface effects on
  magnetic phase transitions},\ }\href
  {https://doi.org/10.1103/PhysRevB.9.2194} {\bibfield  {journal} {\bibinfo
  {journal} {Phys. Rev. B}\ }\textbf {\bibinfo {volume} {9}},\ \bibinfo {pages}
  {2194} (\bibinfo {year} {1974})}\BibitemShut {NoStop}%
\bibitem [{\citenamefont {Binder}(1983)}]{Binder1983}%
  \BibitemOpen
  \bibfield  {author} {\bibinfo {author} {\bibfnamefont {K.}~\bibnamefont
  {Binder}},\ }\bibfield  {title} {\bibinfo {title} {$phase$ $transitions$
  $and$ $critical$ $phenomena$, edited by c. domb and j. l. lebowitz},\
  }\href@noop {} {\bibfield  {journal} {\bibinfo  {journal} {London:
  Academic)}\ }\textbf {\bibinfo {volume} {8}} (\bibinfo {year}
  {1983})}\BibitemShut {NoStop}%
\bibitem [{\citenamefont {Deng}\ \emph {et~al.}(2005)\citenamefont {Deng},
  \citenamefont {Bl\"ote},\ and\ \citenamefont {Nightingale}}]{Deng2005}%
  \BibitemOpen
  \bibfield  {author} {\bibinfo {author} {\bibfnamefont {Y.}~\bibnamefont
  {Deng}}, \bibinfo {author} {\bibfnamefont {H.~W.~J.}\ \bibnamefont
  {Bl\"ote}},\ and\ \bibinfo {author} {\bibfnamefont {M.~P.}\ \bibnamefont
  {Nightingale}},\ }\bibfield  {title} {\bibinfo {title} {Surface and bulk
  transitions in three-dimensional $\mathrm{O}(n)$ models},\ }\href
  {https://doi.org/10.1103/PhysRevE.72.016128} {\bibfield  {journal} {\bibinfo
  {journal} {Phys. Rev. E}\ }\textbf {\bibinfo {volume} {72}},\ \bibinfo
  {pages} {016128} (\bibinfo {year} {2005})}\BibitemShut {NoStop}%
\bibitem [{\citenamefont {Metlitski}(2022)}]{Metlitski2020}%
  \BibitemOpen
  \bibfield  {author} {\bibinfo {author} {\bibfnamefont {M.~A.}\ \bibnamefont
  {Metlitski}},\ }\bibfield  {title} {\bibinfo {title} {{Boundary criticality
  of the O(N) model in d = 3 critically revisited}},\ }\href
  {https://doi.org/10.21468/SciPostPhys.12.4.131} {\bibfield  {journal}
  {\bibinfo  {journal} {SciPost Phys.}\ }\textbf {\bibinfo {volume} {12}},\
  \bibinfo {pages} {131} (\bibinfo {year} {2022})}\BibitemShut {NoStop}%
\bibitem [{\citenamefont {Parisen~Toldin}(2021)}]{ParisenToldin2021}%
  \BibitemOpen
  \bibfield  {author} {\bibinfo {author} {\bibfnamefont {F.}~\bibnamefont
  {Parisen~Toldin}},\ }\bibfield  {title} {\bibinfo {title} {Boundary critical
  behavior of the three-dimensional heisenberg universality class},\ }\href
  {https://doi.org/10.1103/PhysRevLett.126.135701} {\bibfield  {journal}
  {\bibinfo  {journal} {Phys. Rev. Lett.}\ }\textbf {\bibinfo {volume} {126}},\
  \bibinfo {pages} {135701} (\bibinfo {year} {2021})}\BibitemShut {NoStop}%
\bibitem [{\citenamefont {Hu}\ \emph {et~al.}(2021)\citenamefont {Hu},
  \citenamefont {Deng},\ and\ \citenamefont {Lv}}]{Hu2021}%
  \BibitemOpen
  \bibfield  {author} {\bibinfo {author} {\bibfnamefont {M.}~\bibnamefont
  {Hu}}, \bibinfo {author} {\bibfnamefont {Y.}~\bibnamefont {Deng}},\ and\
  \bibinfo {author} {\bibfnamefont {J.-P.}\ \bibnamefont {Lv}},\ }\bibfield
  {title} {\bibinfo {title} {Extraordinary-log surface phase transition in the
  three-dimensional $xy$ model},\ }\href
  {https://doi.org/10.1103/PhysRevLett.127.120603} {\bibfield  {journal}
  {\bibinfo  {journal} {Phys. Rev. Lett.}\ }\textbf {\bibinfo {volume} {127}},\
  \bibinfo {pages} {120603} (\bibinfo {year} {2021})}\BibitemShut {NoStop}%
\bibitem [{\citenamefont {Parisen~Toldin}\ and\ \citenamefont
  {Metlitski}(2022)}]{ParisenToldin2022}%
  \BibitemOpen
  \bibfield  {author} {\bibinfo {author} {\bibfnamefont {F.}~\bibnamefont
  {Parisen~Toldin}}\ and\ \bibinfo {author} {\bibfnamefont {M.~A.}\
  \bibnamefont {Metlitski}},\ }\bibfield  {title} {\bibinfo {title} {Boundary
  criticality of the 3d o($n$) model: From normal to extraordinary},\ }\href
  {https://doi.org/10.1103/PhysRevLett.128.215701} {\bibfield  {journal}
  {\bibinfo  {journal} {Phys. Rev. Lett.}\ }\textbf {\bibinfo {volume} {128}},\
  \bibinfo {pages} {215701} (\bibinfo {year} {2022})}\BibitemShut {NoStop}%
\bibitem [{\citenamefont {Padayasi}\ \emph {et~al.}(2022)\citenamefont
  {Padayasi}, \citenamefont {Krishnan}, \citenamefont {Metlitski},
  \citenamefont {Gruzberg},\ and\ \citenamefont {Meineri}}]{meineri2022}%
  \BibitemOpen
  \bibfield  {author} {\bibinfo {author} {\bibfnamefont {J.}~\bibnamefont
  {Padayasi}}, \bibinfo {author} {\bibfnamefont {A.}~\bibnamefont {Krishnan}},
  \bibinfo {author} {\bibfnamefont {M.~A.}\ \bibnamefont {Metlitski}}, \bibinfo
  {author} {\bibfnamefont {I.~A.}\ \bibnamefont {Gruzberg}},\ and\ \bibinfo
  {author} {\bibfnamefont {M.}~\bibnamefont {Meineri}},\ }\bibfield  {title}
  {\bibinfo {title} {{The extraordinary boundary transition in the 3d O(N)
  model via conformal bootstrap}},\ }\href
  {https://doi.org/10.21468/SciPostPhys.12.6.190} {\bibfield  {journal}
  {\bibinfo  {journal} {SciPost Phys.}\ }\textbf {\bibinfo {volume} {12}},\
  \bibinfo {pages} {190} (\bibinfo {year} {2022})}\BibitemShut {NoStop}%
\bibitem [{\citenamefont {Zou}\ \emph {et~al.}(2022)\citenamefont {Zou},
  \citenamefont {Liu},\ and\ \citenamefont {Guo}}]{Zou2022}%
  \BibitemOpen
  \bibfield  {author} {\bibinfo {author} {\bibfnamefont {X.}~\bibnamefont
  {Zou}}, \bibinfo {author} {\bibfnamefont {S.}~\bibnamefont {Liu}},\ and\
  \bibinfo {author} {\bibfnamefont {W.}~\bibnamefont {Guo}},\ }\bibfield
  {title} {\bibinfo {title} {Surface critical properties of the
  three-dimensional clock model},\ }\href
  {https://doi.org/10.1103/PhysRevB.106.064420} {\bibfield  {journal} {\bibinfo
   {journal} {Phys. Rev. B}\ }\textbf {\bibinfo {volume} {106}},\ \bibinfo
  {pages} {064420} (\bibinfo {year} {2022})}\BibitemShut {NoStop}%
\bibitem [{\citenamefont {Gu}\ and\ \citenamefont {Wen}(2009)}]{XiaoGang2009}%
  \BibitemOpen
  \bibfield  {author} {\bibinfo {author} {\bibfnamefont {Z.-C.}\ \bibnamefont
  {Gu}}\ and\ \bibinfo {author} {\bibfnamefont {X.-G.}\ \bibnamefont {Wen}},\
  }\bibfield  {title} {\bibinfo {title} {Tensor-entanglement-filtering
  renormalization approach and symmetry-protected topological order},\ }\href
  {https://doi.org/10.1103/PhysRevB.80.155131} {\bibfield  {journal} {\bibinfo
  {journal} {Phys. Rev. B}\ }\textbf {\bibinfo {volume} {80}},\ \bibinfo
  {pages} {155131} (\bibinfo {year} {2009})}\BibitemShut {NoStop}%
\bibitem [{\citenamefont {Pollmann}\ \emph
  {et~al.}(2010{\natexlab{a}})\citenamefont {Pollmann}, \citenamefont {Turner},
  \citenamefont {Berg},\ and\ \citenamefont {Oshikawa}}]{Frank2010}%
  \BibitemOpen
  \bibfield  {author} {\bibinfo {author} {\bibfnamefont {F.}~\bibnamefont
  {Pollmann}}, \bibinfo {author} {\bibfnamefont {A.~M.}\ \bibnamefont
  {Turner}}, \bibinfo {author} {\bibfnamefont {E.}~\bibnamefont {Berg}},\ and\
  \bibinfo {author} {\bibfnamefont {M.}~\bibnamefont {Oshikawa}},\ }\bibfield
  {title} {\bibinfo {title} {Entanglement spectrum of a topological phase in
  one dimension},\ }\href {https://doi.org/10.1103/PhysRevB.81.064439}
  {\bibfield  {journal} {\bibinfo  {journal} {Phys. Rev. B}\ }\textbf {\bibinfo
  {volume} {81}},\ \bibinfo {pages} {064439} (\bibinfo {year}
  {2010}{\natexlab{a}})}\BibitemShut {NoStop}%
\bibitem [{\citenamefont {Chen}\ \emph {et~al.}(2012)\citenamefont {Chen},
  \citenamefont {Gu}, \citenamefont {Liu},\ and\ \citenamefont
  {Wen}}]{XiaoGang2012}%
  \BibitemOpen
  \bibfield  {author} {\bibinfo {author} {\bibfnamefont {X.}~\bibnamefont
  {Chen}}, \bibinfo {author} {\bibfnamefont {Z.-C.}\ \bibnamefont {Gu}},
  \bibinfo {author} {\bibfnamefont {Z.-X.}\ \bibnamefont {Liu}},\ and\ \bibinfo
  {author} {\bibfnamefont {X.-G.}\ \bibnamefont {Wen}},\ }\bibfield  {title}
  {\bibinfo {title} {Symmetry-protected topological orders in interacting
  bosonic systems},\ }\href {https://doi.org/10.1126/science.1227224}
  {\bibfield  {journal} {\bibinfo  {journal} {Science}\ }\textbf {\bibinfo
  {volume} {338}},\ \bibinfo {pages} {1604} (\bibinfo {year} {2012})},\ \Eprint
  {https://arxiv.org/abs/https://www.science.org/doi/pdf/10.1126/science.1227224}
  {https://www.science.org/doi/pdf/10.1126/science.1227224} \BibitemShut
  {NoStop}%
\bibitem [{\citenamefont {Haldane}(1983)}]{Haldane}%
  \BibitemOpen
  \bibfield  {author} {\bibinfo {author} {\bibfnamefont {F.~D.~M.}\
  \bibnamefont {Haldane}},\ }\bibfield  {title} {\bibinfo {title} {Nonlinear
  field theory of large-spin heisenberg antiferromagnets: Semiclassically
  quantized solitons of the one-dimensional easy-axis n\'eel state},\ }\href
  {https://doi.org/10.1103/PhysRevLett.50.1153} {\bibfield  {journal} {\bibinfo
   {journal} {Phys. Rev. Lett.}\ }\textbf {\bibinfo {volume} {50}},\ \bibinfo
  {pages} {1153} (\bibinfo {year} {1983})}\BibitemShut {NoStop}%
\bibitem [{\citenamefont {Affleck}\ \emph {et~al.}(1987)\citenamefont
  {Affleck}, \citenamefont {Kennedy}, \citenamefont {Lieb},\ and\ \citenamefont
  {Tasaki}}]{AKLT}%
  \BibitemOpen
  \bibfield  {author} {\bibinfo {author} {\bibfnamefont {I.}~\bibnamefont
  {Affleck}}, \bibinfo {author} {\bibfnamefont {T.}~\bibnamefont {Kennedy}},
  \bibinfo {author} {\bibfnamefont {E.~H.}\ \bibnamefont {Lieb}},\ and\
  \bibinfo {author} {\bibfnamefont {H.}~\bibnamefont {Tasaki}},\ }\bibfield
  {title} {\bibinfo {title} {Rigorous results on valence-bond ground states in
  antiferromagnets},\ }\href {https://doi.org/10.1103/PhysRevLett.59.799}
  {\bibfield  {journal} {\bibinfo  {journal} {Phys. Rev. Lett.}\ }\textbf
  {\bibinfo {volume} {59}},\ \bibinfo {pages} {799} (\bibinfo {year}
  {1987})}\BibitemShut {NoStop}%
\bibitem [{\citenamefont {Zhang}\ and\ \citenamefont {Wang}(2017)}]{Zhang2017}%
  \BibitemOpen
  \bibfield  {author} {\bibinfo {author} {\bibfnamefont {L.}~\bibnamefont
  {Zhang}}\ and\ \bibinfo {author} {\bibfnamefont {F.}~\bibnamefont {Wang}},\
  }\bibfield  {title} {\bibinfo {title} {Unconventional surface critical
  behavior induced by a quantum phase transition from the two-dimensional
  affleck-kennedy-lieb-tasaki phase to a n\'eel-ordered phase},\ }\href
  {https://doi.org/10.1103/PhysRevLett.118.087201} {\bibfield  {journal}
  {\bibinfo  {journal} {Phys. Rev. Lett.}\ }\textbf {\bibinfo {volume} {118}},\
  \bibinfo {pages} {087201} (\bibinfo {year} {2017})}\BibitemShut {NoStop}%
\bibitem [{\citenamefont {Zhu}\ \emph {et~al.}(2021{\natexlab{a}})\citenamefont
  {Zhu}, \citenamefont {Ding}, \citenamefont {Zhang},\ and\ \citenamefont
  {Guo}}]{Zhu2021}%
  \BibitemOpen
  \bibfield  {author} {\bibinfo {author} {\bibfnamefont {W.}~\bibnamefont
  {Zhu}}, \bibinfo {author} {\bibfnamefont {C.}~\bibnamefont {Ding}}, \bibinfo
  {author} {\bibfnamefont {L.}~\bibnamefont {Zhang}},\ and\ \bibinfo {author}
  {\bibfnamefont {W.}~\bibnamefont {Guo}},\ }\bibfield  {title} {\bibinfo
  {title} {Surface critical behavior of coupled haldane chains},\ }\href
  {https://doi.org/10.1103/PhysRevB.103.024412} {\bibfield  {journal} {\bibinfo
   {journal} {Phys. Rev. B}\ }\textbf {\bibinfo {volume} {103}},\ \bibinfo
  {pages} {024412} (\bibinfo {year} {2021}{\natexlab{a}})}\BibitemShut
  {NoStop}%
\bibitem [{\citenamefont {Wang}\ \emph {et~al.}(2022)\citenamefont {Wang},
  \citenamefont {Zhang},\ and\ \citenamefont {Guo}}]{Wang2022}%
  \BibitemOpen
  \bibfield  {author} {\bibinfo {author} {\bibfnamefont {Z.}~\bibnamefont
  {Wang}}, \bibinfo {author} {\bibfnamefont {F.}~\bibnamefont {Zhang}},\ and\
  \bibinfo {author} {\bibfnamefont {W.}~\bibnamefont {Guo}},\ }\bibfield
  {title} {\bibinfo {title} {Bulk and surface critical behavior of a quantum
  heisenberg antiferromagnet on two-dimensional coupled diagonal ladders},\
  }\href {https://doi.org/10.1103/PhysRevB.106.134407} {\bibfield  {journal}
  {\bibinfo  {journal} {Phys. Rev. B}\ }\textbf {\bibinfo {volume} {106}},\
  \bibinfo {pages} {134407} (\bibinfo {year} {2022})}\BibitemShut {NoStop}%
\bibitem [{\citenamefont {Ding}\ \emph {et~al.}(2018)\citenamefont {Ding},
  \citenamefont {Zhang},\ and\ \citenamefont {Guo}}]{Ding2018}%
  \BibitemOpen
  \bibfield  {author} {\bibinfo {author} {\bibfnamefont {C.}~\bibnamefont
  {Ding}}, \bibinfo {author} {\bibfnamefont {L.}~\bibnamefont {Zhang}},\ and\
  \bibinfo {author} {\bibfnamefont {W.}~\bibnamefont {Guo}},\ }\bibfield
  {title} {\bibinfo {title} {Engineering surface critical behavior of
  ($2+1$)-dimensional o(3) quantum critical points},\ }\href
  {https://doi.org/10.1103/PhysRevLett.120.235701} {\bibfield  {journal}
  {\bibinfo  {journal} {Phys. Rev. Lett.}\ }\textbf {\bibinfo {volume} {120}},\
  \bibinfo {pages} {235701} (\bibinfo {year} {2018})}\BibitemShut {NoStop}%
\bibitem [{\citenamefont {Weber}\ \emph {et~al.}(2018)\citenamefont {Weber},
  \citenamefont {Parisen~Toldin},\ and\ \citenamefont {Wessel}}]{Weber2018}%
  \BibitemOpen
  \bibfield  {author} {\bibinfo {author} {\bibfnamefont {L.}~\bibnamefont
  {Weber}}, \bibinfo {author} {\bibfnamefont {F.}~\bibnamefont
  {Parisen~Toldin}},\ and\ \bibinfo {author} {\bibfnamefont {S.}~\bibnamefont
  {Wessel}},\ }\bibfield  {title} {\bibinfo {title} {Nonordinary edge
  criticality of two-dimensional quantum critical magnets},\ }\href
  {https://doi.org/10.1103/PhysRevB.98.140403} {\bibfield  {journal} {\bibinfo
  {journal} {Phys. Rev. B}\ }\textbf {\bibinfo {volume} {98}},\ \bibinfo
  {pages} {140403} (\bibinfo {year} {2018})}\BibitemShut {NoStop}%
\bibitem [{\citenamefont {Ding}\ \emph {et~al.}(2021)\citenamefont {Ding},
  \citenamefont {Zhu}, \citenamefont {Guo},\ and\ \citenamefont
  {Zhang}}]{Ding2021}%
  \BibitemOpen
  \bibfield  {author} {\bibinfo {author} {\bibfnamefont {C.}~\bibnamefont
  {Ding}}, \bibinfo {author} {\bibfnamefont {W.}~\bibnamefont {Zhu}}, \bibinfo
  {author} {\bibfnamefont {W.}~\bibnamefont {Guo}},\ and\ \bibinfo {author}
  {\bibfnamefont {L.}~\bibnamefont {Zhang}},\ }\href@noop {} {\bibinfo {title}
  {Special transition and extraordinary phase on the surface of a
  (2+1)-dimensional quantum heisenberg antiferromagnet}} (\bibinfo {year}
  {2021}),\ \Eprint {https://arxiv.org/abs/2110.04762} {arXiv:2110.04762
  [cond-mat.str-el]} \BibitemShut {NoStop}%
\bibitem [{\citenamefont {Lieb}\ \emph {et~al.}(1961)\citenamefont {Lieb},
  \citenamefont {Schultz},\ and\ \citenamefont {Mattis}}]{LIEB1961}%
  \BibitemOpen
  \bibfield  {author} {\bibinfo {author} {\bibfnamefont {E.}~\bibnamefont
  {Lieb}}, \bibinfo {author} {\bibfnamefont {T.}~\bibnamefont {Schultz}},\ and\
  \bibinfo {author} {\bibfnamefont {D.}~\bibnamefont {Mattis}},\ }\bibfield
  {title} {\bibinfo {title} {Two soluble models of an antiferromagnetic
  chain},\ }\href
  {https://doi.org/https://doi.org/10.1016/0003-4916(61)90115-4} {\bibfield
  {journal} {\bibinfo  {journal} {Annals of Physics}\ }\textbf {\bibinfo
  {volume} {16}},\ \bibinfo {pages} {407} (\bibinfo {year} {1961})}\BibitemShut
  {NoStop}%
\bibitem [{\citenamefont {Kim}\ \emph {et~al.}(2000)\citenamefont {Kim},
  \citenamefont {F\'ath}, \citenamefont {S\'olyom},\ and\ \citenamefont
  {Scalapino}}]{diagonal2000}%
  \BibitemOpen
  \bibfield  {author} {\bibinfo {author} {\bibfnamefont {E.~H.}\ \bibnamefont
  {Kim}}, \bibinfo {author} {\bibfnamefont {G.}~\bibnamefont {F\'ath}},
  \bibinfo {author} {\bibfnamefont {J.}~\bibnamefont {S\'olyom}},\ and\
  \bibinfo {author} {\bibfnamefont {D.~J.}\ \bibnamefont {Scalapino}},\
  }\bibfield  {title} {\bibinfo {title} {Phase transitions between
  topologically distinct gapped phases in isotropic spin ladders},\ }\href
  {https://doi.org/10.1103/PhysRevB.62.14965} {\bibfield  {journal} {\bibinfo
  {journal} {Phys. Rev. B}\ }\textbf {\bibinfo {volume} {62}},\ \bibinfo
  {pages} {14965} (\bibinfo {year} {2000})}\BibitemShut {NoStop}%
\bibitem [{\citenamefont {Buyers}\ \emph {et~al.}(1986)\citenamefont {Buyers},
  \citenamefont {Morra}, \citenamefont {Armstrong}, \citenamefont {Hogan},
  \citenamefont {Gerlach}, \citenamefont {Hirakawa},\ and\ \citenamefont
  {K.}}]{Buyers1986}%
  \BibitemOpen
  \bibfield  {author} {\bibinfo {author} {\bibfnamefont {W.~J.~L.}\
  \bibnamefont {Buyers}}, \bibinfo {author} {\bibfnamefont {R.~M.}\
  \bibnamefont {Morra}}, \bibinfo {author} {\bibfnamefont {R.~L.}\ \bibnamefont
  {Armstrong}}, \bibinfo {author} {\bibfnamefont {M.~J.}\ \bibnamefont
  {Hogan}}, \bibinfo {author} {\bibfnamefont {P.}~\bibnamefont {Gerlach}},
  \bibinfo {author} {\bibnamefont {Hirakawa}},\ and\ \bibinfo {author}
  {\bibnamefont {K.}},\ }\bibfield  {title} {\bibinfo {title} {Experimental
  evidence for the haldane gap in a spin-1 nearly isotropic, antiferromagnetic
  chain},\ }\href {https://doi.org/10.1103/PhysRevLett.56.371} {\bibfield
  {journal} {\bibinfo  {journal} {Phys. Rev. Lett.}\ }\textbf {\bibinfo
  {volume} {56}},\ \bibinfo {pages} {371} (\bibinfo {year} {1986})}\BibitemShut
  {NoStop}%
\bibitem [{\citenamefont {Dagotto}(1999)}]{ElbioDagotto1999}%
  \BibitemOpen
  \bibfield  {author} {\bibinfo {author} {\bibfnamefont {E.}~\bibnamefont
  {Dagotto}},\ }\bibfield  {title} {\bibinfo {title} {Experiments on ladders
  reveal a complex interplay between a spin-gapped normal state and
  superconductivity},\ }\href {https://doi.org/10.1088/0034-4885/62/11/202}
  {\bibfield  {journal} {\bibinfo  {journal} {Reports on Progress in Physics}\
  }\textbf {\bibinfo {volume} {62}},\ \bibinfo {pages} {1525} (\bibinfo {year}
  {1999})}\BibitemShut {NoStop}%
\bibitem [{\citenamefont {Wierschem}\ and\ \citenamefont
  {Sengupta}(2014)}]{Wierschem2014}%
  \BibitemOpen
  \bibfield  {author} {\bibinfo {author} {\bibfnamefont {K.}~\bibnamefont
  {Wierschem}}\ and\ \bibinfo {author} {\bibfnamefont {P.}~\bibnamefont
  {Sengupta}},\ }\bibfield  {title} {\bibinfo {title} {Quenching the haldane
  gap in spin-1 heisenberg antiferromagnets},\ }\href
  {https://doi.org/10.1103/PhysRevLett.112.247203} {\bibfield  {journal}
  {\bibinfo  {journal} {Phys. Rev. Lett.}\ }\textbf {\bibinfo {volume} {112}},\
  \bibinfo {pages} {247203} (\bibinfo {year} {2014})}\BibitemShut {NoStop}%
\bibitem [{\citenamefont {Sandvik}\ and\ \citenamefont
  {Kurkij\"arvi}(1991)}]{Sandviksusc1991}%
  \BibitemOpen
  \bibfield  {author} {\bibinfo {author} {\bibfnamefont {A.~W.}\ \bibnamefont
  {Sandvik}}\ and\ \bibinfo {author} {\bibfnamefont {J.}~\bibnamefont
  {Kurkij\"arvi}},\ }\bibfield  {title} {\bibinfo {title} {Quantum monte carlo
  simulation method for spin systems},\ }\href
  {https://doi.org/10.1103/PhysRevB.43.5950} {\bibfield  {journal} {\bibinfo
  {journal} {Phys. Rev. B}\ }\textbf {\bibinfo {volume} {43}},\ \bibinfo
  {pages} {5950} (\bibinfo {year} {1991})}\BibitemShut {NoStop}%
\bibitem [{\citenamefont {Sandvik}(1999)}]{Sandvik1999}%
  \BibitemOpen
  \bibfield  {author} {\bibinfo {author} {\bibfnamefont {A.~W.}\ \bibnamefont
  {Sandvik}},\ }\bibfield  {title} {\bibinfo {title} {Stochastic series
  expansion method with operator-loop update},\ }\href
  {https://doi.org/10.1103/PhysRevB.59.R14157} {\bibfield  {journal} {\bibinfo
  {journal} {Phys. Rev. B}\ }\textbf {\bibinfo {volume} {59}},\ \bibinfo
  {pages} {R14157} (\bibinfo {year} {1999})}\BibitemShut {NoStop}%
\bibitem [{\citenamefont {Binder}(1981)}]{Binder1981}%
  \BibitemOpen
  \bibfield  {author} {\bibinfo {author} {\bibfnamefont {K.}~\bibnamefont
  {Binder}},\ }\bibfield  {title} {\bibinfo {title} {Critical properties from
  monte carlo coarse graining and renormalization},\ }\href
  {https://doi.org/10.1103/PhysRevLett.47.693} {\bibfield  {journal} {\bibinfo
  {journal} {Phys. Rev. Lett.}\ }\textbf {\bibinfo {volume} {47}},\ \bibinfo
  {pages} {693} (\bibinfo {year} {1981})}\BibitemShut {NoStop}%
\bibitem [{\citenamefont {Binder}\ and\ \citenamefont
  {Landau}(1984)}]{Binder1984}%
  \BibitemOpen
  \bibfield  {author} {\bibinfo {author} {\bibfnamefont {K.}~\bibnamefont
  {Binder}}\ and\ \bibinfo {author} {\bibfnamefont {D.~P.}\ \bibnamefont
  {Landau}},\ }\bibfield  {title} {\bibinfo {title} {Finite-size scaling at
  first-order phase transitions},\ }\href
  {https://doi.org/10.1103/PhysRevB.30.1477} {\bibfield  {journal} {\bibinfo
  {journal} {Phys. Rev. B}\ }\textbf {\bibinfo {volume} {30}},\ \bibinfo
  {pages} {1477} (\bibinfo {year} {1984})}\BibitemShut {NoStop}%
\bibitem [{\citenamefont {Pollock}\ and\ \citenamefont
  {Ceperley}(1987)}]{Pollock1987}%
  \BibitemOpen
  \bibfield  {author} {\bibinfo {author} {\bibfnamefont {E.~L.}\ \bibnamefont
  {Pollock}}\ and\ \bibinfo {author} {\bibfnamefont {D.~M.}\ \bibnamefont
  {Ceperley}},\ }\bibfield  {title} {\bibinfo {title} {Path-integral
  computation of superfluid densities},\ }\href
  {https://doi.org/10.1103/PhysRevB.36.8343} {\bibfield  {journal} {\bibinfo
  {journal} {Phys. Rev. B}\ }\textbf {\bibinfo {volume} {36}},\ \bibinfo
  {pages} {8343} (\bibinfo {year} {1987})}\BibitemShut {NoStop}%
\bibitem [{\citenamefont {Sandvik}(2010)}]{SandvikAIP}%
  \BibitemOpen
  \bibfield  {author} {\bibinfo {author} {\bibfnamefont {A.~W.}\ \bibnamefont
  {Sandvik}},\ }\bibfield  {title} {\bibinfo {title} {Computational studies of
  quantum spin systems},\ }\href {https://doi.org/10.1063/1.3518900} {\bibfield
   {journal} {\bibinfo  {journal} {AIP Conference Proceedings}\ }\textbf
  {\bibinfo {volume} {1297}},\ \bibinfo {pages} {135} (\bibinfo {year}
  {2010})}\BibitemShut {NoStop}%
\bibitem [{\citenamefont {Fisher}\ \emph {et~al.}(1989)\citenamefont {Fisher},
  \citenamefont {Weichman}, \citenamefont {Grinstein},\ and\ \citenamefont
  {Fisher}}]{Fisher1989}%
  \BibitemOpen
  \bibfield  {author} {\bibinfo {author} {\bibfnamefont {M.~P.~A.}\
  \bibnamefont {Fisher}}, \bibinfo {author} {\bibfnamefont {P.~B.}\
  \bibnamefont {Weichman}}, \bibinfo {author} {\bibfnamefont {G.}~\bibnamefont
  {Grinstein}},\ and\ \bibinfo {author} {\bibfnamefont {D.~S.}\ \bibnamefont
  {Fisher}},\ }\bibfield  {title} {\bibinfo {title} {Boson localization and the
  superfluid-insulator transition},\ }\href
  {https://doi.org/10.1103/PhysRevB.40.546} {\bibfield  {journal} {\bibinfo
  {journal} {Phys. Rev. B}\ }\textbf {\bibinfo {volume} {40}},\ \bibinfo
  {pages} {546} (\bibinfo {year} {1989})}\BibitemShut {NoStop}%
\bibitem [{\citenamefont {Shao}\ \emph {et~al.}(2016)\citenamefont {Shao},
  \citenamefont {Guo},\ and\ \citenamefont {Sandvik}}]{shao_science}%
  \BibitemOpen
  \bibfield  {author} {\bibinfo {author} {\bibfnamefont {H.}~\bibnamefont
  {Shao}}, \bibinfo {author} {\bibfnamefont {W.}~\bibnamefont {Guo}},\ and\
  \bibinfo {author} {\bibfnamefont {A.~W.}\ \bibnamefont {Sandvik}},\
  }\bibfield  {title} {\bibinfo {title} {Quantum criticality with two length
  scales},\ }\href {https://doi.org/10.1126/science.aad5007} {\bibfield
  {journal} {\bibinfo  {journal} {Science}\ }\textbf {\bibinfo {volume}
  {352}},\ \bibinfo {pages} {213} (\bibinfo {year} {2016})},\ \Eprint
  {https://arxiv.org/abs/https://www.science.org/doi/pdf/10.1126/science.aad5007}
  {https://www.science.org/doi/pdf/10.1126/science.aad5007} \BibitemShut
  {NoStop}%
\bibitem [{\citenamefont {Guida}\ and\ \citenamefont
  {Zinn-Justin}(1998)}]{Guida1998}%
  \BibitemOpen
  \bibfield  {author} {\bibinfo {author} {\bibfnamefont {R.}~\bibnamefont
  {Guida}}\ and\ \bibinfo {author} {\bibfnamefont {J.}~\bibnamefont
  {Zinn-Justin}},\ }\bibfield  {title} {\bibinfo {title} {Critical exponents of
  {theN}-vector model},\ }\href {https://doi.org/10.1088/0305-4470/31/40/006}
  {\bibfield  {journal} {\bibinfo  {journal} {Journal of Physics A:
  Mathematical and General}\ }\textbf {\bibinfo {volume} {31}},\ \bibinfo
  {pages} {8103} (\bibinfo {year} {1998})}\BibitemShut {NoStop}%
\bibitem [{\citenamefont {Hasenbusch}\ and\ \citenamefont
  {Vicari}(2011)}]{Hasenbusch2011}%
  \BibitemOpen
  \bibfield  {author} {\bibinfo {author} {\bibfnamefont {M.}~\bibnamefont
  {Hasenbusch}}\ and\ \bibinfo {author} {\bibfnamefont {E.}~\bibnamefont
  {Vicari}},\ }\bibfield  {title} {\bibinfo {title} {Anisotropic perturbations
  in three-dimensional o($n$)-symmetric vector models},\ }\href
  {https://doi.org/10.1103/PhysRevB.84.125136} {\bibfield  {journal} {\bibinfo
  {journal} {Phys. Rev. B}\ }\textbf {\bibinfo {volume} {84}},\ \bibinfo
  {pages} {125136} (\bibinfo {year} {2011})}\BibitemShut {NoStop}%
\bibitem [{\citenamefont {Matsumoto}\ \emph {et~al.}(2001)\citenamefont
  {Matsumoto}, \citenamefont {Yasuda}, \citenamefont {Todo},\ and\
  \citenamefont {Takayama}}]{Matsumoto2001}%
  \BibitemOpen
  \bibfield  {author} {\bibinfo {author} {\bibfnamefont {M.}~\bibnamefont
  {Matsumoto}}, \bibinfo {author} {\bibfnamefont {C.}~\bibnamefont {Yasuda}},
  \bibinfo {author} {\bibfnamefont {S.}~\bibnamefont {Todo}},\ and\ \bibinfo
  {author} {\bibfnamefont {H.}~\bibnamefont {Takayama}},\ }\bibfield  {title}
  {\bibinfo {title} {Ground-state phase diagram of quantum heisenberg
  antiferromagnets on the anisotropic dimerized square lattice},\ }\href
  {https://doi.org/10.1103/PhysRevB.65.014407} {\bibfield  {journal} {\bibinfo
  {journal} {Phys. Rev. B}\ }\textbf {\bibinfo {volume} {65}},\ \bibinfo
  {pages} {014407} (\bibinfo {year} {2001})}\BibitemShut {NoStop}%
\bibitem [{\citenamefont {Affleck}\ and\ \citenamefont
  {Haldane}(1987)}]{Affleck}%
  \BibitemOpen
  \bibfield  {author} {\bibinfo {author} {\bibfnamefont {I.}~\bibnamefont
  {Affleck}}\ and\ \bibinfo {author} {\bibfnamefont {F.~D.~M.}\ \bibnamefont
  {Haldane}},\ }\bibfield  {title} {\bibinfo {title} {Critical theory of
  quantum spin chains},\ }\href {https://doi.org/10.1103/PhysRevB.36.5291}
  {\bibfield  {journal} {\bibinfo  {journal} {Phys. Rev. B}\ }\textbf {\bibinfo
  {volume} {36}},\ \bibinfo {pages} {5291} (\bibinfo {year}
  {1987})}\BibitemShut {NoStop}%
\bibitem [{\citenamefont {Pollmann}\ \emph
  {et~al.}(2010{\natexlab{b}})\citenamefont {Pollmann}, \citenamefont {Turner},
  \citenamefont {Berg},\ and\ \citenamefont {Oshikawa}}]{Pollmann2010}%
  \BibitemOpen
  \bibfield  {author} {\bibinfo {author} {\bibfnamefont {F.}~\bibnamefont
  {Pollmann}}, \bibinfo {author} {\bibfnamefont {A.~M.}\ \bibnamefont
  {Turner}}, \bibinfo {author} {\bibfnamefont {E.}~\bibnamefont {Berg}},\ and\
  \bibinfo {author} {\bibfnamefont {M.}~\bibnamefont {Oshikawa}},\ }\bibfield
  {title} {\bibinfo {title} {Entanglement spectrum of a topological phase in
  one dimension},\ }\href {https://doi.org/10.1103/PhysRevB.81.064439}
  {\bibfield  {journal} {\bibinfo  {journal} {Phys. Rev. B}\ }\textbf {\bibinfo
  {volume} {81}},\ \bibinfo {pages} {064439} (\bibinfo {year}
  {2010}{\natexlab{b}})}\BibitemShut {NoStop}%
\bibitem [{\citenamefont {den Nijs}\ and\ \citenamefont
  {Rommelse}(1989)}]{Nijs1989}%
  \BibitemOpen
  \bibfield  {author} {\bibinfo {author} {\bibfnamefont {M.}~\bibnamefont {den
  Nijs}}\ and\ \bibinfo {author} {\bibfnamefont {K.}~\bibnamefont {Rommelse}},\
  }\bibfield  {title} {\bibinfo {title} {Preroughening transitions in crystal
  surfaces and valence-bond phases in quantum spin chains},\ }\href
  {https://doi.org/10.1103/PhysRevB.40.4709} {\bibfield  {journal} {\bibinfo
  {journal} {Phys. Rev. B}\ }\textbf {\bibinfo {volume} {40}},\ \bibinfo
  {pages} {4709} (\bibinfo {year} {1989})}\BibitemShut {NoStop}%
\bibitem [{\citenamefont {Anfuso}\ and\ \citenamefont
  {Rosch}(2007)}]{Anfuso2007}%
  \BibitemOpen
  \bibfield  {author} {\bibinfo {author} {\bibfnamefont {F.}~\bibnamefont
  {Anfuso}}\ and\ \bibinfo {author} {\bibfnamefont {A.}~\bibnamefont {Rosch}},\
  }\bibfield  {title} {\bibinfo {title} {Fragility of string orders},\ }\href
  {https://doi.org/10.1103/PhysRevB.76.085124} {\bibfield  {journal} {\bibinfo
  {journal} {Phys. Rev. B}\ }\textbf {\bibinfo {volume} {76}},\ \bibinfo
  {pages} {085124} (\bibinfo {year} {2007})}\BibitemShut {NoStop}%
\bibitem [{\citenamefont {Wen}(2017)}]{wen2017}%
  \BibitemOpen
  \bibfield  {author} {\bibinfo {author} {\bibfnamefont {X.-G.}\ \bibnamefont
  {Wen}},\ }\bibfield  {title} {\bibinfo {title} {Colloquium: Zoo of
  quantum-topological phases of matter},\ }\href
  {https://doi.org/10.1103/RevModPhys.89.041004} {\bibfield  {journal}
  {\bibinfo  {journal} {Rev. Mod. Phys.}\ }\textbf {\bibinfo {volume} {89}},\
  \bibinfo {pages} {041004} (\bibinfo {year} {2017})}\BibitemShut {NoStop}%
\bibitem [{\citenamefont {Diehl}(1986)}]{Diehl}%
  \BibitemOpen
  \bibfield  {author} {\bibinfo {author} {\bibfnamefont {H.~W.}\ \bibnamefont
  {Diehl}},\ }\href@noop {} {\emph {\bibinfo {title} {Phase Transitions and
  Critical Phenomena}}},\ edited by\ \bibinfo {editor} {\bibfnamefont
  {C.}~\bibnamefont {Domb}}\ and\ \bibinfo {editor} {\bibfnamefont {J.~L.}\
  \bibnamefont {Lebowitz}},\ Vol.~\bibinfo {volume} {10}\ (\bibinfo
  {publisher} {Academic},\ \bibinfo {address} {London},\ \bibinfo {year}
  {1986})\ pp.\ \bibinfo {pages} {75--267}\BibitemShut {NoStop}%
\bibitem [{\citenamefont {Jian}\ \emph {et~al.}(2021)\citenamefont {Jian},
  \citenamefont {Xu}, \citenamefont {Wu},\ and\ \citenamefont
  {Xu}}]{JianCM2021}%
  \BibitemOpen
  \bibfield  {author} {\bibinfo {author} {\bibfnamefont {C.-M.}\ \bibnamefont
  {Jian}}, \bibinfo {author} {\bibfnamefont {Y.}~\bibnamefont {Xu}}, \bibinfo
  {author} {\bibfnamefont {X.-C.}\ \bibnamefont {Wu}},\ and\ \bibinfo {author}
  {\bibfnamefont {C.}~\bibnamefont {Xu}},\ }\bibfield  {title} {\bibinfo
  {title} {{Continuous N\'{e}el-VBS Quantum Phase Transition in Non-Local
  one-dimensional systems with SO(3) Symmetry}},\ }\href
  {https://doi.org/10.21468/SciPostPhys.10.2.033} {\bibfield  {journal}
  {\bibinfo  {journal} {SciPost Phys.}\ }\textbf {\bibinfo {volume} {10}},\
  \bibinfo {pages} {33} (\bibinfo {year} {2021})}\BibitemShut {NoStop}%
\bibitem [{\citenamefont {Zhu}\ \emph {et~al.}(2021{\natexlab{b}})\citenamefont
  {Zhu}, \citenamefont {Ding}, \citenamefont {Zhang},\ and\ \citenamefont
  {Guo}}]{Zhu2021b}%
  \BibitemOpen
  \bibfield  {author} {\bibinfo {author} {\bibfnamefont {W.}~\bibnamefont
  {Zhu}}, \bibinfo {author} {\bibfnamefont {C.}~\bibnamefont {Ding}}, \bibinfo
  {author} {\bibfnamefont {L.}~\bibnamefont {Zhang}},\ and\ \bibinfo {author}
  {\bibfnamefont {W.}~\bibnamefont {Guo}},\ }\href@noop {} {\bibinfo {title}
  {Exotic surface behaviors induced by geometrical settings of the
  two-dimensional dimerized quantum xxz model}} (\bibinfo {year}
  {2021}{\natexlab{b}}),\ \Eprint {https://arxiv.org/abs/2111.12336}
  {arXiv:2111.12336 [cond-mat.str-el]} \BibitemShut {NoStop}%
\end{thebibliography}%

\end{document}